\documentclass[usenatbib,usegraphicx,useAMS]{mn2e} 

\begin{document}
                 
\title[Dynamical modelling of ESO 566-24]{Dynamical modelling of a
  remarkable four-armed barred spiral galaxy ESO 566-24}    
\author[Rautiainen et al.]
{P. ~Rautiainen$^{1}$\thanks{E-mail: pertti.rautiainen@oulu.fi}
  H. ~Salo$^{1}$ R. ~Buta$^{2}$\\ 
$^1$Department of Physical Sciences, University of Oulu,  P.O. Box
  3000, FIN-90014 Oulun yliopisto, Finland\\
$^2$Department of Physics and Astronomy, University of Alabama, Box
  870324, Tuscaloosa, Alabama 35486, USA}

\date{Received x.x, accepted x.x}

\maketitle

\begin{abstract}
ESO 566-24 is an extraordinary barred galaxy, which has four regularly
spaced spiral arms in blue light images. This type of four-armed
spiral structure, which is rare among the spiral population, is
clearly seen also in near-infrared images, and thus is present in the old
stellar population. We have constructed dynamical models of ESO 566-24
in order to understand the cause of its four-armed structure.
The disk gravitational potential is determined from near-infrared
photometry, and the gas dynamics is modelled as inelastically
colliding particles. The resulting morphology and kinematics with
different assumed pattern speeds, disk vertical thicknesses and dark
halo contributions is compared with observations. Our models can
reproduce the main morphological features of this galaxy: the
four-armed spiral, and the inner and nuclear rings. The pattern speed
of the bar in this galaxy is such that the corotation resonance
is well outside the bar radius, $r_{CR}/r_{bar} = 1.6 \pm 0.3$. The
four-armed spiral resides in the region between inner and outer
4/1-resonances. Also, the main kinematical features, including
bar-induced deviations from circular rotation,  are explained by our
models. The best fit is obtained when the dark halo contribution is
just enough to make the modelled rotation curve match the observed
one. This ``minimum halo'' is rather moderate: luminous matter dominates
the rotation curve within the disk region.
\end{abstract}


\section{Introduction}
\label{introduction}

Barred galaxies constitute the majority of all spiral galaxies. According
to recent near-infrared (NIR) studies, about 70\%
of disk galaxies have a bar \citep{eskridge2000,knapen2000}. Barred
galaxies often have rings or pseudorings in their disks with very
distinctive characteristics \citep{buta96b}. Depending on
its relative size with respect to the bar, a ring can be
classified as an outer ring (semi-major axis about twice the bar
radius), an inner ring (semi-major axis $\approx$ bar radius) or
a nuclear ring (semi-major axis about 10 \% of the bar
radius, though the scatter is large) \citep{buta93a}. Although all
three types of barred galaxy rings can show enhanced star formation
(see, for example, \citealt{buta98a}),
nuclear rings are most prone to starbursts \citep{maoz96,buta2000a}. In
addition to rings, barred spiral galaxies often show offset
dust lanes on the leading sides of the bar (assuming trailing spiral arms). 

The dynamical behaviour of 
barred galaxies has been modelled in different ways. Some studies are
general, in the sense that no specific barred galaxy is modelled, but
they concentrate instead on issues such as the formation and evolution of
the bar. In most of these studies, the large-scale bars are
suggested to be fast rotators, i.e.\ their corotation resonance (CR) radius
should be near the end of the bar \citep{sellwood81}. This is in
accordance with gas dynamical simulations of the shapes of the leading
offset dust lanes inside the bars \citep{athanassoula92b}. Exceptions
to fast rotating bars are the studies by \citet{combes93} and
\citet{miwa98}; in the former study a short and slow rotating bar forms in a simulation
where the disk was strongly truncated radially,
while in the latter, a slow bar forms as a result of an interaction. The
contribution of a dark halo inside the radius of the optical disk can
have a strong effect on disk dynamics. According to \citet{ostriker73}
a massive halo can inhibit the bar formation, and
\citet{debattista98,debattista2000} have shown simulations where the
interaction between a bar and a halo decelerates strongly the bar
rotation. However, conflicting results have also been published:
\citet{athanassoula2002a} and \citet{athanassoula2002b} show examples
where the halo actually promotes bar formation, and \citet{valenzuela2002}
found a rather low angular momentum transfer rate from the bar to the
dark halo.  

In principle, the pattern speed of a bar can be
determined from observations for a disk with a well-defined
pattern speed \citep{tremaine84}. The Tremaine-Weinberg method
further assumes that the used tracer component follows the continuity
equation, i.e.\ its intensity is proportional to total density. When
these conditions are fulfilled, the pattern speed can be determined
from the observed intensity and the line-of-sight velocity along a strip
parallel to galaxy major axis. In most of the cases, where this method
has been applied, the corotation resonance seems to be near the
end of the bar \citep{kent87,merrifield95,aguerri2003}. Note, however,
that due to its limitations, this method has been successfully applied
mostly to SB0-galaxies. 

Another kinematical method to determine the pattern speed
was presented by \citet{canzian93} and is based on 
observed residuals in the velocity field after subtracting the
axisymmetric velocity component. In the case of a two-armed spiral, the
velocity residuals exhibit an $m=1$ spiral shape inside corotation, whereas
outside corotation, an $m=3$
spiral should appear. This method was used to determine
the pattern speed in NGC 4321 by \citet{sempere95a} and
\citet{canzian97}, who found a rather low value, setting the corotation
radius about 1.6 -- 1.8 times the bar radius. However, the method is
very sensitive to noise, and the construction of the axisymmetric 
velocity component is not trivial. Purcell (1998), who
studied a larger galaxy sample, found that
only in 7 of 27 Sbc galaxies could the expected pattern be seen in
velocity field residuals.  

There are several suggested resonance indicators related to
morphological details of spiral structure \citep{elmegreen89b}. Even
if these are valid, their application to barred galaxies is not
necessarily straightforward: there can be more than one mode in the system.
For example the outer spiral arms can have a lower pattern speed than the
bar \citep{sellwood88}. In such cases the pattern speed derived e.g.\ from
the dust lane morphology of the outer arms would not correspond to
that of the bar.  

The formation of rings in barred galaxies has been explained by the
response of the dissipative gas component to the gravitational torque
of the bar
\citep{schwarz81,combes85,byrd94,buta96b,rautiainen2000}. The shapes
of the rings resemble the shapes of certain closed periodic orbits in
barred potentials \citep{contopoulos89b}. When the
non-axisymmetric perturbation is weak, these rings form very close to
resonance radii.
The usual resonance identifications are such that the outer rings are
located near the outer Lindblad resonance (OLR), the inner rings near
the inner 4/1-resonance and the nuclear rings near the inner Lindblad
resonance (ILR). Thus if the rotation curve of the galaxy is known,
the sizes of the rings can be used to estimate the bar pattern
speed. However, these pattern speed estimates can be misleading when the
bar perturbation is strong, and the resonance distances based on the
linear approximation fail. This is especially the case with nuclear
rings, which are located in a region where the
relative non-axisymmetric forcing is strong.  
Furthermore, both nuclear and outer rings can often be affected by
another mode in addition to the main bar; nuclear rings by a secondary
bar, and outer rings by an outer spiral mode
\citep{rautiainen2000,rautiainen2002}.   

The effect of the strength of the bar is taken into account when
detailed models for individual galaxies are being constructed. The
adopted approach has varied from fitting an analytical mass model to
observations \citep{duval83} to determining the mass model from
near-infrared observations \citep{lindblad96a,lindblad96b}. In some models, the
non-axisymmetric forcing of the bar was not enough to reproduce the
spiral structure with the observed extent, and so an unseen
oval-shaped component was added \citep{england89,ball92}. A spiral
potential corresponding to the NIR-morphology solved this problem more
elegantly \citep{lindblad96a}. The number of successfully modelled
barred galaxies is still quite small, and it is not clear how well
the ranges of parameters such as the bar pattern speed and the
relative strength of the non-axisymmetric perturbation are known. 

We started our modelling project of ringed galaxies with the
normal, but rather weakly barred galaxy IC 4214. From the observations
\citep{buta99b}, we constructed a mass model with a few free
parameters \citep{salo99}. In our best-fitting models both the
morphology and kinematics were well reproduced. Following the
successful modelling of IC 4214, we
extend our studies here by constructing a dynamical model for an
extraordinary barred spiral: the four-armed galaxy ESO 566-24.

\section{A Remarkable Four-Armed Barred Spiral}

ESO 566-24 is a barred spiral galaxy of de Vaucouleurs
revised Hubble type SB(r)b. It was first recognized as 
being an exceptional example of a four-armed ringed barred 
spiral galaxy during early stages in the production of the Catalog
of Southern Ringed Galaxies \citep[CSRG, ][]{buta95a}. High quality
optical CCD images in the $U$, $B$, and $I$ passbands,
obtained in follow-up work by \citet{buta91b}, reveal
an extremely regular four-armed pattern of normal spiral arms,
as well as star formation in the inner ring and in a very small
previously unknown nuclear ring. These images also revealed a
faint, partial outer ring-like pattern that envelopes the
four-armed spiral pattern \citep{buta98b}. Although rings are not rare
in barred galaxies, regular four-armed outer spirals are very rare
and, if seen at all, appear in a partial manner, as the secondary
spiral arcs and two main outer arms seen in NGC 1433
\citep{buta86b,buta2001a}.  

\citet{buta98b} showed that the four-armed structure of ESO 566-24
is clearly seen in a NIR $H$-band image. Thus, the pattern is also
present in the old stellar population. A recent study of
about 200 galaxies in different wavebands by \citet{eskridge2002}
demonstrated that the NIR morphology of galaxies is usually not as
dramatically different from the optical morphology as is sometimes stated
\citep{block99a}, with multiple-arm and flocculent morphology also being 
seen in the $H$-band. However, in most of the cases, two
arms dominate, and the extra arms are weaker and usually form as a
bifurcation of the main arms. This is not the case with
ESO 566-24: the four arms are of about same strength and length.

In this paper, we use numerical models to find the parameter range
(pattern speed, mass-to-light ratio) which can reproduce the regular
four-armed morphology of ESO 566-24. We study how resonances are
located with respect to the observed structures such as rings, and we
also identify orbits supporting these structures. Finally, we compare
the spectacular morphology of ESO 566-24 with more partial patterns
seen in other barred galaxies.  

\begin{figure*} 
\resizebox{\hsize}{!}{\includegraphics{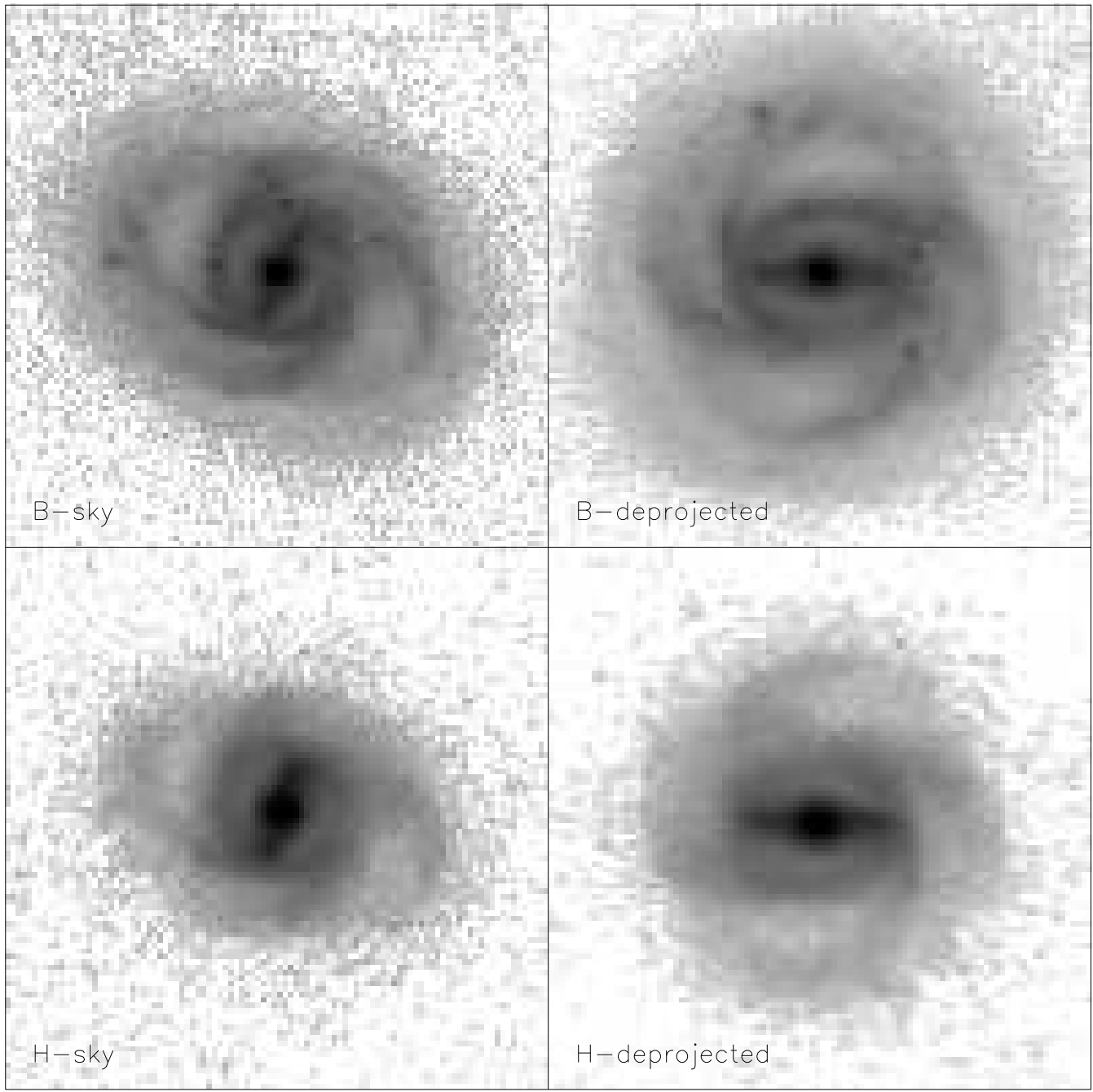}}  \hfill
\caption{B- and H-band images of ESO 566-24. The images on the left
  side show the galaxy as it is oriented on the sky (foreground stars
  have been removed) and the images on the right side have been first
  deprojected and then rotated so that the bar is horizontal. The
  width of the frames is 120$\arcsec$.}
\label{esoimage} 
\end{figure*}

\section{Methods}
\label{methods}

The method used by \citet{salo99} in modelling the early-type, weakly-barred
spiral galaxy IC 4214 is used here for modelling ESO 566$-$24.
The gas morphology is modelled by inelastically colliding test
particles, which move in a rigidly-rotating gravitational potential.
We estimate the mass
distribution from NIR $H$-band photometry, assuming a constant
mass-to-light ratio throughout the disk. Before
deprojecting the galaxy image, a bulge--disk decomposition is done, and the
resulting bulge component is subtracted from the image. Then the disk
surface brightness (and surface density) distribution is approximated
using a Fourier decomposition   

\begin{eqnarray}
\Sigma(r,\theta) &=& \Sigma_\mathrm{0}(r)\left\{ 1 +
\sum_{m=1}^{\infty}{A_\mathrm{m}(r)} \cos {\left[ m \left( \theta -
    \theta_\mathrm{m} \right) \right]}
\right\} \nonumber\\
&=& \Sigma_0(r)+
\sum_{m=1}^{\infty} \Sigma_m(r,\theta), 
\label{decompose} 
\end{eqnarray}

\noindent where $r$ and $\theta$ are the polar coordinates,
$\Sigma(r,\theta)$ the disk surface density, $\Sigma_\mathrm{0}$ the
axisymmetric surface density, and $A_\mathrm{m}(r)$ and
$\theta_\mathrm{m}(r)$ are the Fourier amplitude and phase
angles. For our analysis, we used even components $m = 0$ -- $8$. This
introduces smoothing of small scale structure such as star forming
regions but has little or no impact on the basic background stellar
potential. A similar procedure was adopted by
\citet{kranz2001}. From the density components
$\Sigma_m$, we then calculate the potential components $\Phi_m$

\begin{equation}
\Phi_m (r,\theta,z=0) = -G \int_{0}^{\infty} r' dr' \int_{0}^{2\pi}
\Sigma_m(r', \theta') g(\Delta r) d\theta', 
\label{poteq}
\end{equation}

\noindent where $(\Delta r)^2= {r'}^2+r^2-2rr'cos(\theta'-\theta)$ and the
convolution function

\begin{equation}
g(\Delta r) = \int_{-\infty}^{\infty} \varrho_z(z) ((\Delta r)^2+z^2)^{-1
/2} dz 
\label{convolution}
\end{equation}

\noindent is used for calculating the
effect of the vertical density profile $\varrho(z)$ (normalized to unity when
integrated over the vertical direction). 
In \citet{salo99} the gravity softening was used as a measure of finite disk
thickness. In this study we use an exponential vertical distribution,

\begin{equation}
\varrho (z) = {1 \over {2 h_z}} exp (-|z|/h_z)
\label{expthick}
\end{equation}

\noindent which has a similar effect. A more detailed description of
the potential calculation can be found in \citet{laurikainen2002b}. We
have tried scale heights in the range $1/12$ -- $1/2$ times the radial
exponential scale length $h_r$ of the disk, corresponding to the
change from a very thin disk to an unrealistically thick one; for the
morphological type of ESO 566-24, $h_z/h_r \approx 1/6$ is expected
\citep{degrijs97,degrijs98}. Some tests were made with variable scale
height, making the disk thicker in the bar region.    

The non-axisymmetric perturbation of the potential is turned on
gradually during four bar rotation periods. This procedure was adopted
to smoothly ``guide'' particles to realistic orbits, thus
trying to minimize the effect of transient phenomena related to abrupt
changes in the potential, e.g.\  such as a large number of gas particles 
ending up on crossing orbits.

The gas component is modelled as a two-dimensional disk consisting of 
20,000 inelastically colliding test particles. In each collision, the
relative velocity component parallel with the line joining the
particle centres is reversed and multiplied by the coefficient of
restitution. In each time step, only one collision per particle is
allowed. The particle motion is integrated using leap-frog integration. 
The time-step is chosen so that at the radius of the smallest
morphological detail, the nuclear ring, one circular rotation period
is about 50 steps. In addition to gas particles, we have also made
simulations with 200,000 non-colliding ``stellar'' particles with
different initial velocity dispersions. For more details on the
simulation code, see \citet{salo99}.  

We use a modified version of the bulge model presented in
\citet{buta98b}. We take the average radial profile of the bulge,
assume the bulge to be axisymmetric and calculate its gravitational
potential using the algorithm described in \citet{kent86}. As a
possible halo component we use an isothermal sphere with a constant core
radius, which has a rotation curve 

\begin{equation}
v_h(r) = v_\infty \sqrt{r^2 \over {r^2+r_c^2}},
\label{halo}
\end{equation}

\noindent where $v_\infty$ is the asymptotic velocity at infinity and
$r_c$ is the core radius. 

\section{Observed characteristics of ESO 566-24}

Because the observations and data reduction are described in
\citet{buta98b}, we here just give the main morphological and
kinematical characteristics of ESO 566-24, concentrating on details
which are useful for comparing simulations with observations.

ESO 566-24 has a clear bar component in both visual and NIR-images. The
bar is considerably stronger than in IC
4214. It is difficult to judge what would be the best method to
determine the bar length. Applying several methods to $H$-band images:
e.g., axis ratios of isophotes and local amplitude minima of $m=2$ and $m=4$ Fourier
components of disk surface brightness, we estimate the bar radius
to be $r_{bar} = 18\arcsec \pm 2\arcsec$, or about 1.6
times the exponential scale length of the disk, $h_r \approx 11 \farcs
3 $. This is in reasonable agreement with the value adopted in
\citet{buta98b}, $r_{bar} = 16 \farcs 7 $. \citet{athanassoula2002a}
use seven different methods to determine the bar length in N-body
simulations. However, we found that the complicated morphology of ESO
566-24 made many of these unsuitable (e.g.\ their $m=2$
phase angle criterion would give the bar radius in the middle of the
spiral pattern). Colour index maps, e.g.\ $B-I$, show that there are dust
lanes on the leading sides of the bar \citep{buta91b}.  

The bar is surrounded by an inner ring, which appears circular in the
plane of the sky but deprojects to an ellipse parallel with the bar
major axis for the adopted orientation parameters: position angle $PA
= 73 \degr$ and inclination $i = 43 \degr$ \citep{buta98b}. 
Parallel alignment is a normal characteristic of barred galaxy inner
rings \citep{buta95a}. The deprojected
semimajor axis radius of the inner ring is about $19 \arcsec$. There
is also a nuclear ring with a semimajor axis radius of about $4
\arcsec$. The four bright spiral arms extend between $20\arcsec$ and
$40\arcsec$. The arms start from the inner ring, two from its major
axis and the other two from the minor axis. Both the inner ring and
the four spiral arms can be seen also in the NIR $H$-band image. There is
also a single outer spiral arm or a partial outer ring, that is not
connected to the four-armed spiral \citep{buta98b}. Fig.~\ref{esoimage} 
shows $B$- and $H$-band images (as observed in the sky and deprojected) 
of ESO 566-24. 

The rotation curve inferred from the velocity field (see
Fig.~\ref{omecomp}) shows signs of probable non-circular motions: in
the inner parts it rises steeply, then it has a short flat part at
about 120 $\mbox{km s}^{-1}$. Outside this area, it rises steeply
again to the peak value, after which it quickly drops to the constant
level of about 190 $\mbox{km s}^{-1}$. It is probable that the
peculiar inner rotation curve is at least partly caused by
non-circular motions induced by the bar, since axisymmetric rotation
curves constructed from any realistic combinations of bulge, disk and
halo cannot produce such a structure. Also the zero velocity curve of
the velocity field implies the presence of non-circular
motion. However, the effect is not as strong as in IC 4214 because in
ESO 566$-$24, the orientation of the bar is almost perpendicular to
the line of nodes, and hence little of the radial component of the
velocity vector projects to the line of sight \citep[see
  also][]{buta88}. 

\section{Modelling}
\label{models}

We have constructed three major sets of models: series D without a halo
component, series M with a moderate halo and series H with a dominating 
halo (see Table~\ref{modelinfo}). In all of these model series, we have
varied several parameters such as the bar pattern speed and the disk 
thickness. 

\begin{figure*} 
\resizebox{\hsize}{!}{\includegraphics{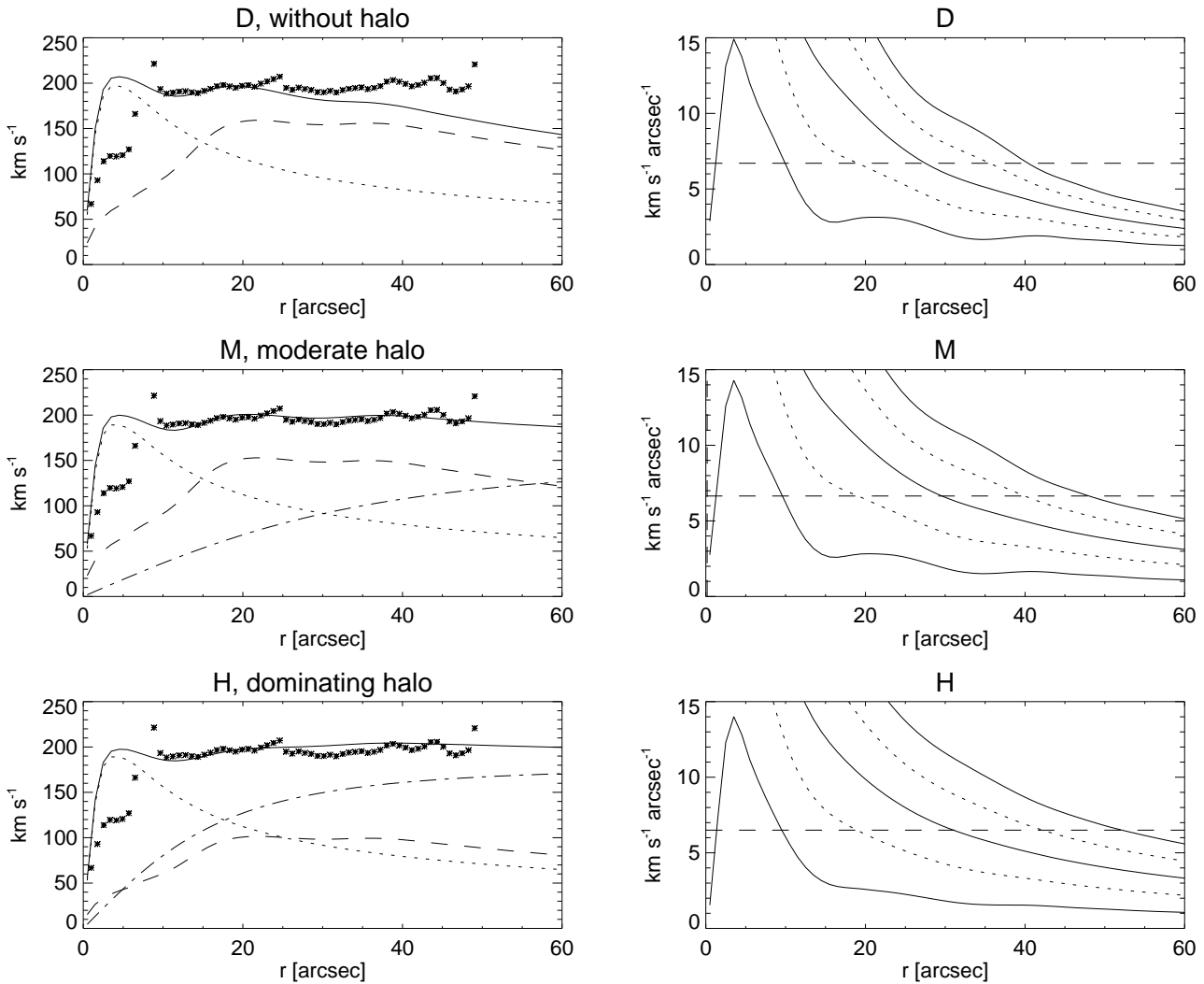}}  \hfill
\caption{Comparison between three models with different mass
  distributions (see Table~\ref{modelinfo}). The frames on the left side
  show the rotation curves, total rotation curve indicated as
  a continuous line, bulge as a dotted line, disk as a dashed line and
  the halo as a dash-dotted line. The asterisk symbols show circular
  velocities determined from the observed velocity field (see
  text). The frames on the right side show $\Omega$ (thick
  continuous line), $\Omega \pm \kappa/2$ (thin continuous lines) and
  $\Omega \pm \kappa/4$ (dotted lines) The dashed horizontal line
  shows the pattern  speed at which the semimajor axis of the inner ring
  ($19 \arcsec$) coincides with the inner 4/1-resonance. The $M/L$ of
  the bulge was chosen in each case so that the rotation amplitude at
  $20 \arcsec$  corresponds with the observed one.}   
\label{omecomp} 
\end{figure*}

The left side frames of Fig.~\ref{omecomp} compare the rotation
curves of models of series D (no halo), M (moderate
halo) and H (dominating halo) with a similar disk thickness ($h_z/h_r
= 1/6$). Also shown are
the circular velocity points determined from the observed velocity
field using the method presented by \citet{warner73}. If we
adopt the definition of a maximal disk (or disk+bulge) by
\citet{sackett97}: a disk is maximal if its rotation speed at radius
$r_{2.2} = 2.2 \ h_r$ (usually close to the peak of the disk
rotation curve) is 75--95\% of the total circular speed at this
radius, then our moderate halo series M resides near the lower
limit of maximal disks. Perhaps a better name would be a minimum halo
model: only such an amount of halo was included that is needed to make
the outer curve as flat as the observed rotation curve.

\begin{table}
\caption{The main parameters for the three model families D, M and
  H. The H-band mass-luminosity ratios of the bulge $(M/L)_\mathrm{B}$ and
  the disk $(M/L)_\mathrm{D}$ are given in solar units, assuming
  $M_{H\sun} = 3.37$ and the distance being 45 Mpc. Also listed are
  the parameters determining the halo rotation curve:
  velocity in infinity $v_\infty$ in $\mathrm{km s}^{-1}$ and core
  radius $r_c$ in arcseconds.}
\begin{tabular}{lcccc}
\hline Model family & $(M/L)_\mathrm{B}$ & $(M/L)_\mathrm{D}$ &
$v_\infty$  & $r_c$\\
\hline
D & 1.3 & 1.3 & & \\
M & 1.2 & 1.2 & 152 & 40.0\\
H & 1.2 & 0.54 & 180 & 20.0\\ 
\hline
\end{tabular}
\label{modelinfo}
\end{table}

The right side frames of Fig.~\ref{omecomp} show the corresponding
``frequency diagrams'', i.e.\ curves showing $\Omega$, $\Omega \pm
\kappa/2$ and $\Omega \pm \kappa/4$, where $\Omega$ is the circular
frequency and $\kappa$ is the epicyclic frequency. 
The inner parts of the models have
very similar rotation curves, and thus with the same pattern speed, the
inner resonance distances (ILRs and inner 4/1) would be about the
same. On the other hand, the outer resonance radii are different. For
example, the distance between the inner and outer 4/1-resonances is
larger in models with a halo than in models without it. 
Throughout this article we use $\mbox{km s}^{-1} \mbox{arcsec}^{-1}$
as a unit of pattern speed. We denote models of each mass model series
with the pattern speed: for example model D6.7 is a series D model
with pattern speed $\Omega_b = 6.7 \ \mbox{km s}^{-1}
\mbox{arcsec}^{-1}$. At the adopted distance of 45 Mpc, $1 \ \mbox{km s}^{-1}
\mbox{arcsec}^{-1}$ corresponds to about $4.6 \ \mbox{km s}^{-1}
\mbox{kpc}^{-1}$ and the length of a simulation is about 2.7 Gyr
(about 14 bar rotation periods in our best-fitting case).  

\subsection{The effect of model parameters}
\label{params}

\subsubsection{The effect of disk thickness}
\label{diskthick}

\begin{figure*} 
\resizebox{\hsize}{!}{\includegraphics{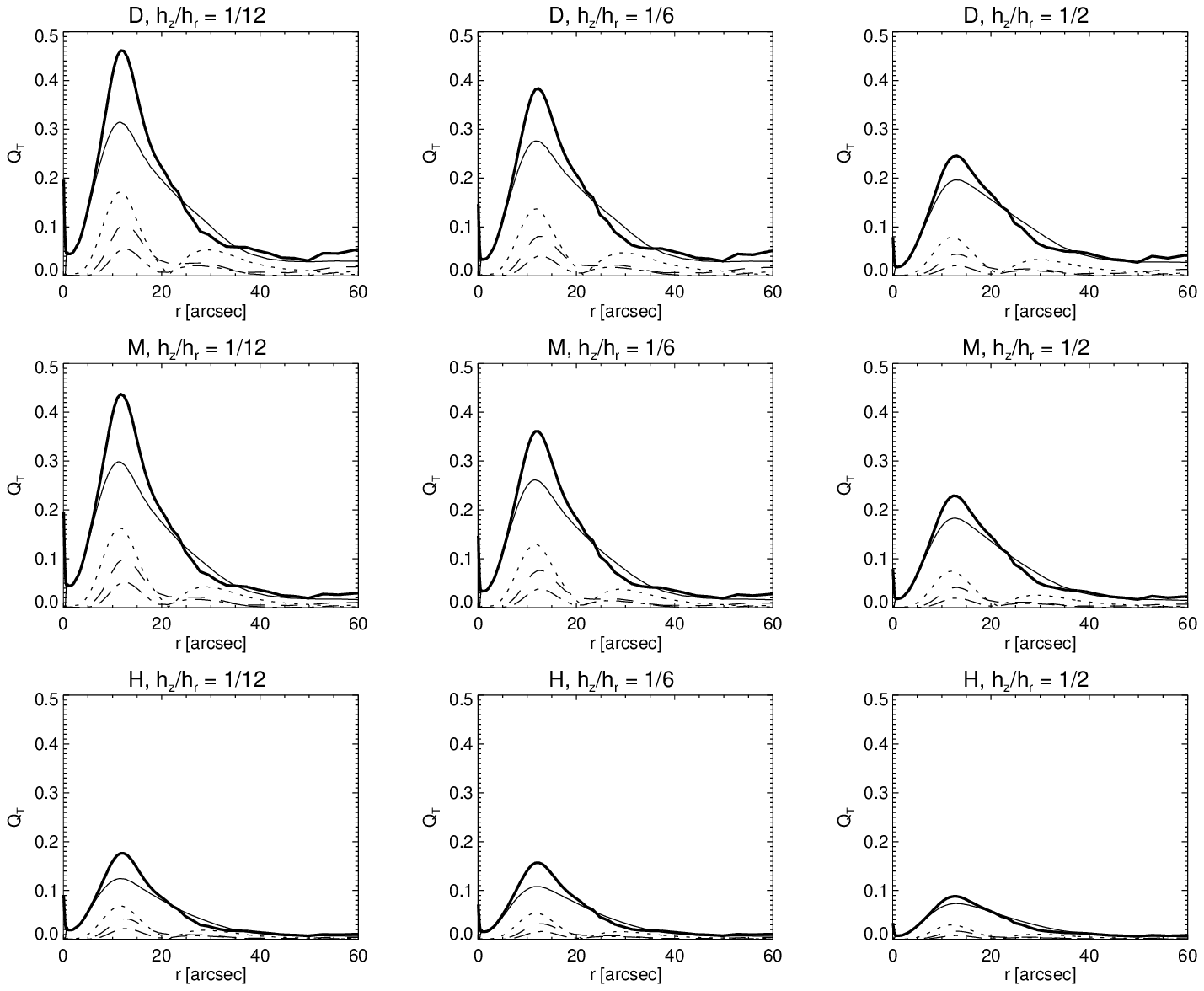}}  \hfill
\caption{The relative tangential force in
  models without a halo (top row), with a moderate halo (middle row)
  and with a strong halo (bottom row). The thick continuous line shows the
  maximum, whereas the m=2
  component is indicated with a thin continuous line, m=4 component with a
  dotted line, m=6 with a dashed line and m=8 with a dot-dashed
  line. Note that owing to phase differences, the maximum is less than
  the sum of the amplitudes of the different components.}
\label{forcecompa} 
\end{figure*}

One of the main model parameters is the disk thickness. We have made
simulations with a range of the disk scale height from 1/12 to 1/2
of the disk scale length, which should cover the plausible range of
values for the disk thickness. In
Fig.~\ref{forcecompa} we show the Fourier components of the tangential
force relative to the axisymmetric radial force (total of bulge, disk and
halo) as a function of radius for different mass models.

As can be seen in Fig.~\ref{forcecompa}, the $m=2$
component has one maximum, at about $12 \arcsec$, well inside the
bar. The $m=4$ component has two clear maxima, one coinciding with the
$m=2$ maximum, and the other broader one in about the middle of the
four-armed spiral structure. The $m=6$ and $m=8$ components have also
two maxima, the inner coinciding with the $m=2$ maximum, and the outer
ones being inside the outer $m=4$ maximum. 

As a measure of bar strength, we use the relative
tangential force $Q_T$ \citep{combes81}: 

\begin{equation}
\label{qb}
Q_T (r) = F_T^{max}(r) / <F_R(r)>,
\end{equation}
 
\noindent where $F_T^{max}(r)$ is the maximum of tangential force at a given
radius $r$, and $<F_R(r)>$ is the azimuthally averaged radial force at the
same radius. The maximum of $Q_T(r)$ over radius, or $Q_b$, has been
used as a single measure of bar strength by \citet{buta2001b} and
\citet{laurikainen2002a}. From Fig.~\ref{forcecompa} we can see that
in our basic mass models, the extreme values of the $Q_b$-parameter
vary from 0.1 to 0.47 when going from the thickest disk models of
series H to the thinnest disk models of series D. In each mass model
series, the models with the thinnest disks have about twice as strong
bars as models with thickest disks if measured by $Q_b$. One can also see
from Fig.~\ref{forcecompa} that the effect of disk thickness on $Q_T$
is much weaker in the region of the spiral arms.

\begin{figure*} 
\resizebox{\hsize}{!}{\includegraphics{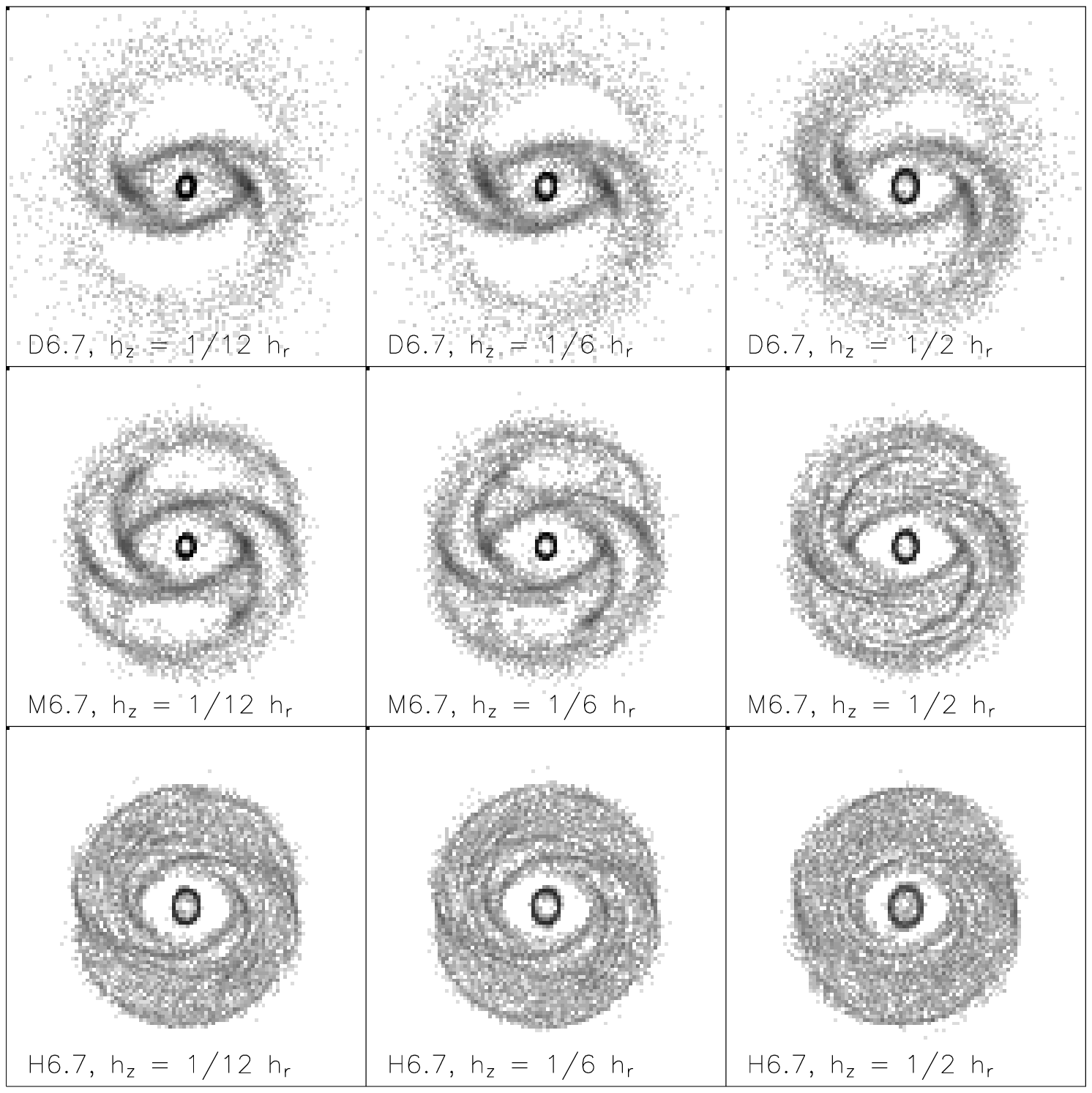}}  \hfill
\caption{The effect of the disk thickness in models from the
  disk-dominated series D (top row), series M with a moderate halo
  (middle row) and the halo-dominated series H (bottom row). Compare
  to Fig.~\ref{forcecompa} showing the strength of the
  non-axisymmetric force field in each case. The width
  of the frames is $120 \arcsec$. The models are shown one bar period
  after the bar has reached its full strength.} 
\label{compthick} 
\end{figure*}

In Fig.~\ref{compthick} we show the effect of disk thickness on the
simulated gas morphology. When comparing with the observed four-armed
spiral of ESO 566-24, we see that in the D-series the fit is best when
the disk is very thick ($h_z/h_r = 1/2$). In series M, the best fit is
obtained with a thinner disk (both $h_z/h_r = 1/12$ and
$h_z/h_r = 1/6$ give a fairly good fit). In series H, the spiral
response to the potential is always too weak. This can be explained by the
weakness of the non-axisymmetric perturbation at the spiral region: it
is less than half of that in series' M and D. We can also see that the
size of the nuclear ring depends on the disk thickness: a thicker disk
leads to a larger ring due to the reduced non-axisymmetric perturbation at its
location, shifting the ring closer to linear ILR. The axial ratio of
the inner ring is not very sensitive to $Q_b$, which is not surprising,
because the location where the maximum of $Q_T$ is reached is well
inside the ring. At the ring region, $Q_T$ varies less than its
maximum value $Q_b$ when compared in models with different disk scale
heights.

\subsubsection{The effect of pattern speed}
\label{pattern}

For each basic mass model, we ran a series of simulations with five
different pattern speeds. The middle pattern speed of each series was
chosen so that the semimajor axis radius of the inner ring coincides 
with the inner 4/1-resonance, and the highest pattern speed so that its
corotation radius is 1.1 times the bar radius. The other values of the
pattern speed were chosen around the middle one with a step size $1.2
\ \mbox{km s}^{-1}\mbox{arcsec}^{-1}$.
 
\begin{figure*} 
\resizebox{\hsize}{!}{\includegraphics{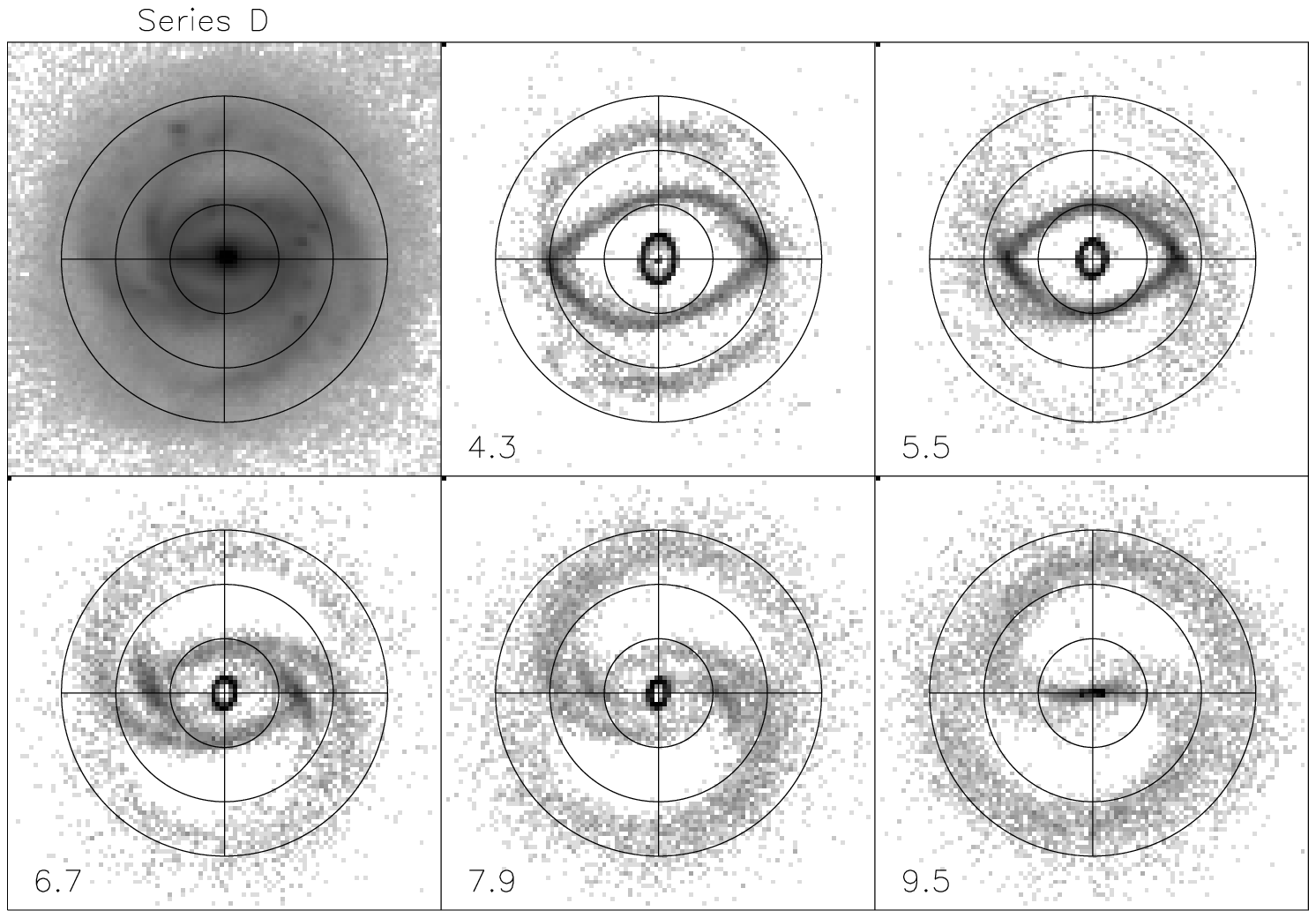}}  \hfill
\caption{Pattern speed series for models without a halo compared with
  a B-band image of ESO 566-24. The circles
  indicate radii of $15$, $30$ and $45\arcsec$. The width of the frames
  is $120 \arcsec$. The models are shown one bar period
  after the bar has reached its full strength.}
\label{disk_omega} 
\end{figure*}

Fig.~\ref{disk_omega} shows the effect of the pattern speed on gas
morphology in series D. When the pattern speed rises,
the nuclear ring becomes smaller. With the highest pattern speed,
$\Omega_b = 9.5$, the inner ring disappears and the nuclear ring
becomes very elongated and parallel with respect to the
bar. Four-armed spiral structure can be seen clearly in models D5.5
and D6.7. Both at higher and lower pattern speeds, the spiral tends to
become two-armed. Several models in this and
other mass model series have weak overdensities of gas particles in the
bar area resembling leading offset dust lanes. These features are
stronger in the early phases of simulations and more or less disappear
during the simulations. These features move closer to the bar when the
pattern speed is increased, but in general they are too vague to be
reliably compared with the observed dust lanes.   

\begin{figure*} 
\resizebox{\hsize}{!}{\includegraphics{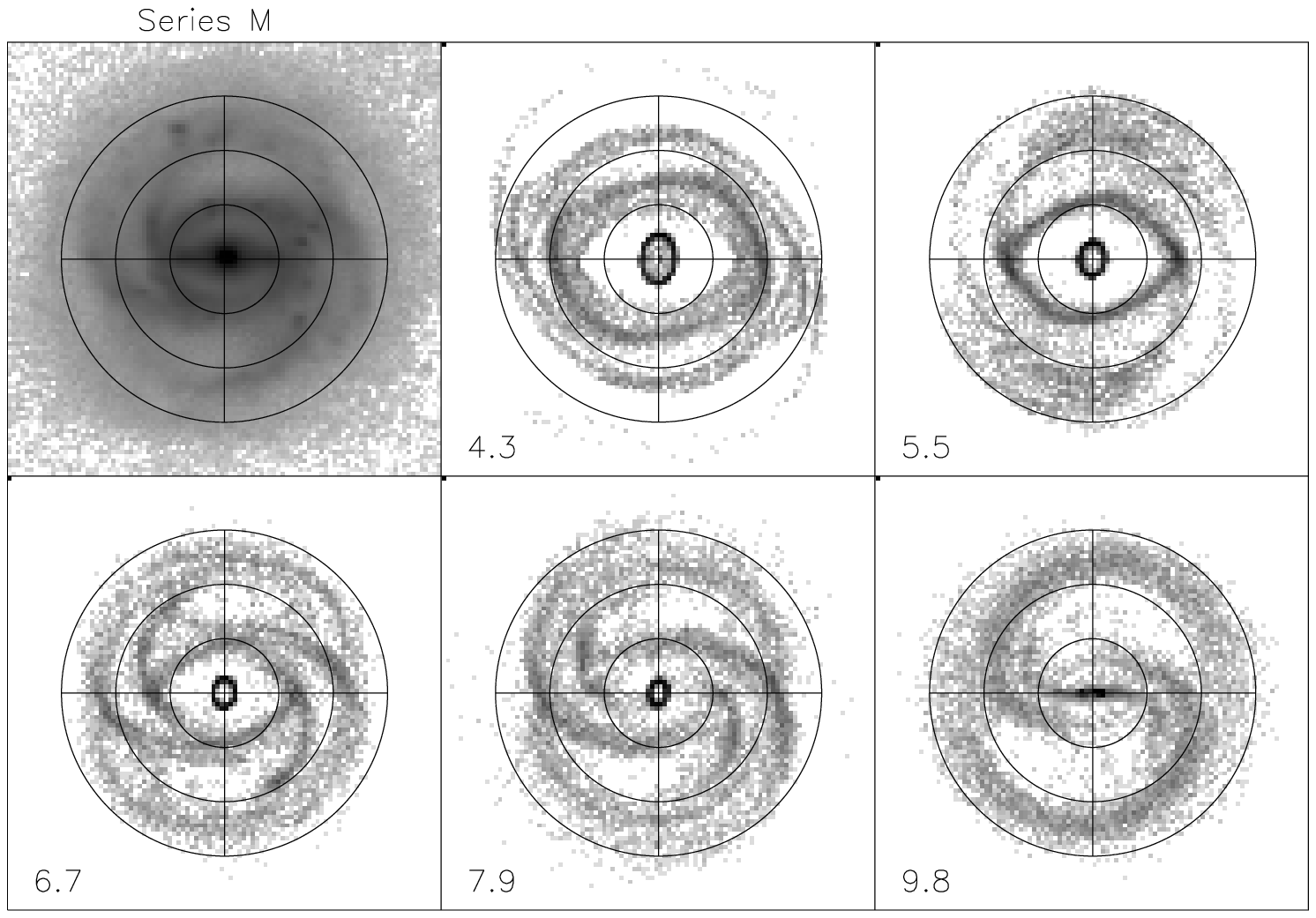}}  \hfill
\caption{Pattern speed series for models with a moderate halo. The circles
  indicate radii of $15$, $30$ and $45 \arcsec$. The width of the frames
  is $120 \arcsec$. The models are shown one bar period
  after the bar has reached its full strength.}
\label{mod_omega} 
\end{figure*}

Fig.~\ref{mod_omega} shows the effect of the pattern speed on the gas
morphology in series M, which has a moderate halo component. The basic trend
is quite similar as in the models without a halo, but the four-armed
spiral appears in a larger pattern speed range (models M5.5 --
M7.9), and is also more pronounced, the arms being longer and sharper.

\begin{figure*} 
\resizebox{\hsize}{!}{\includegraphics{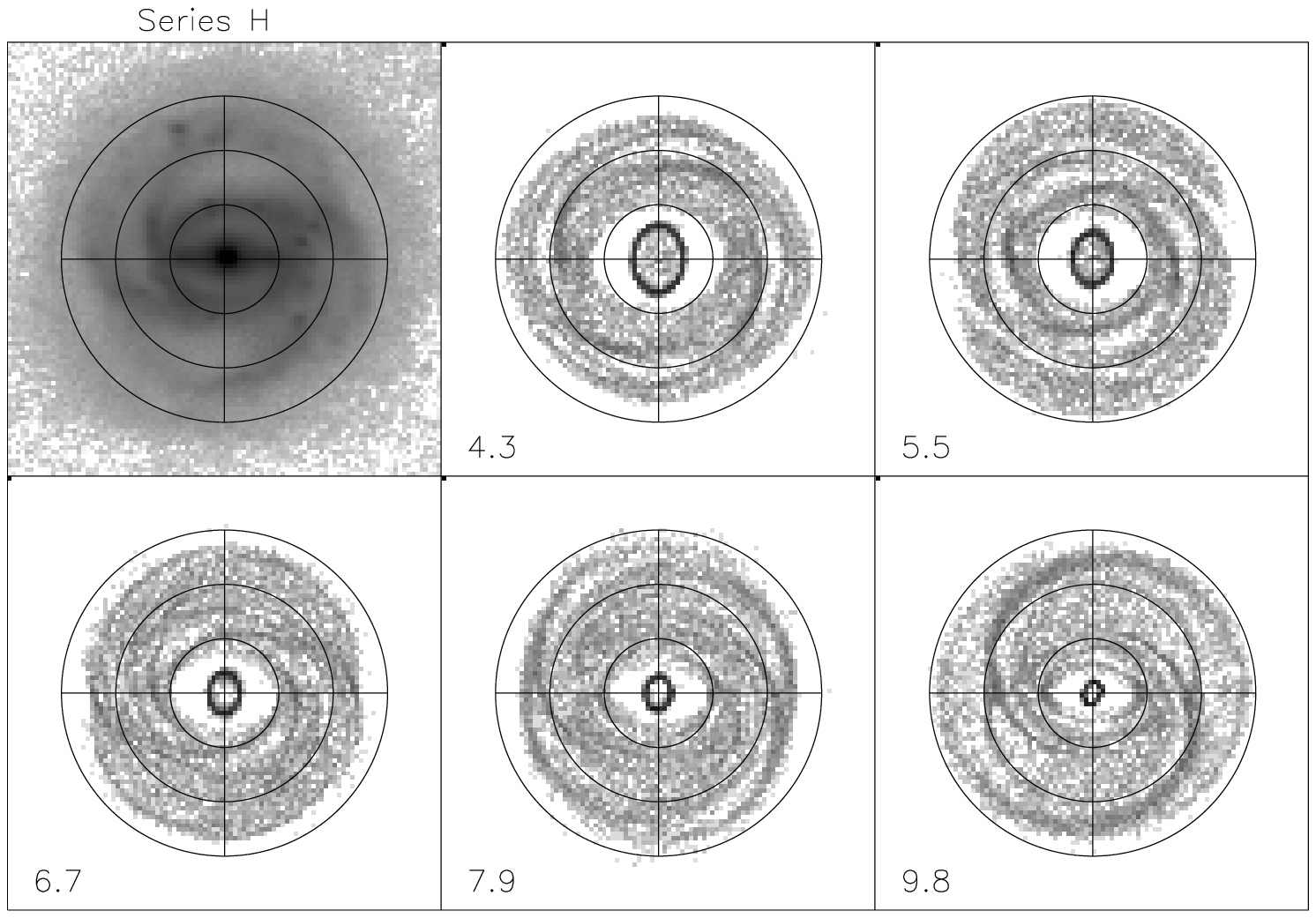}}  \hfill
\caption{Pattern speed series for models with a strong halo. The circles
  indicate radii of $15$, $30$ and $45 \arcsec$. The width of the frames
  is $120 \arcsec$. The models are shown one bar period
  after the bar has reached its full strength.}
\label{halo_omega} 
\end{figure*}

Fig.~\ref{halo_omega} shows the effect of the pattern speed on gas
morphology in series H, with a dominating halo. The
non-axisymmetric force is too weak to produce sharp spiral arms, and thus
only weak spiral features appear, exhibiting a four-armed structure in models
H5.5 -- H7.9.

\subsubsection{Miscellaneous parametric dependencies}

We have also checked the effects of various other model parameters:
the number of Fourier components that were used to calculate the
gravitational potential, the initial gas distribution, the bar turn-on
speed, the collision frequency, and the coefficient of restitution. All
these have only a minor effect when compared to disk thickness and the
pattern speed of the bar. 

The inclusion of Fourier component $m=4$ and the higher ones makes the
four-armed spiral more pronounced. When only $m=0$ and $m=2$
components are included, the response is not a clear $m=4$ spiral, but
resembles model M6.7 with a thick disk. The other mass
models follow similar trends. We have also made simulations where we
include the odd components of the Fourier decomposition. This
causes some asymmetry to the models, especially in the region where
spiral arms emerge from the inner ring, but the effect is small. 

We found that the bar turn-on speed can affect the innermost
morphology: when the bar was turned on too abruptly (obtaining full
strength in one or less bar rotation periods), some of the
nuclear rings collapsed toward the centre or became very elongated and
parallel to the bar. This is probably due to the high amount of
intersecting orbits. When the bar reached its full strength more
gently (in 2--4 bar rotation periods), these gas particles were slowly
``guided'' to orbits perpendicular with respect to the bar. However,
when the pattern speed was high enough, even the gradual bar turn-on
time could not save the nuclear rings, and the short-lived rings were
destroyed before the bar reached its full strength. The reason for
such behaviour is that the number of intersecting orbits increases
when $\Omega_b$ is increased. This was confirmed by simulations using
non-colliding test particles. 

We studied the effect of collision frequency by making selected
simulations with different gas particle sizes. If extreme values are
ignored, the collision frequency affects mainly the sharpness of
features and their formation timescale. However, the main
morphological features are not strongly dependent on the collision
frequency. We also studied the effect of the coefficient of
restitution $\alpha$. Changing $\alpha$ between our standard value,
0.0, and 0.5 has only a minor effect on the morphology. The greatest
difference is in nuclear ring: adopting higher values of $\alpha$
produces a smaller and thicker ring.

\subsection{A search for the best fit}
\label{bestfitting}

We use here the results of the previously presented simulation series
to deduce the probable range of the essential model parameters.

\subsubsection{Morphology}
\label{morphology}

\begin{figure*} 
\resizebox{\hsize}{!}{\includegraphics{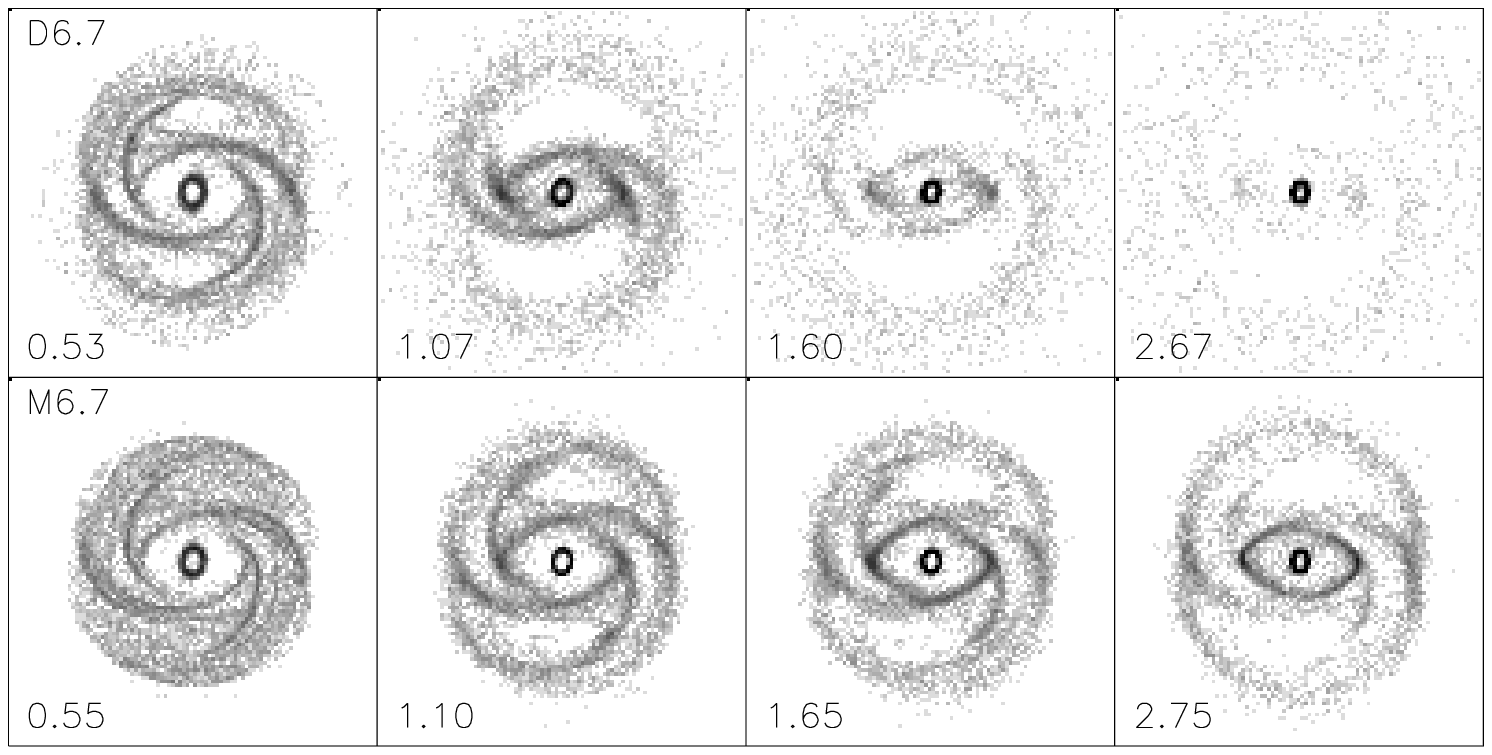}}  \hfill
\caption{The time evolution of two models from series D and M. The
  width of the frames is $120 \arcsec$ and time is shown in
  simulation Gyr. The timesteps of the models are the same, but the
  times differ due to different mass scaling.}
\label{esoevol} 
\end{figure*}

The top row of Fig.~\ref{esoevol} shows the time evolution of
morphologically-selected best-fitting models from the D- and M-series (in
series H the spiral response is always very weak). The models
without a halo, including the one shown, tend to have a problem with
the positions of the four spiral arms: the arms are not equally spaced
in azimuth. Instead of a clear $m=4$ spiral, the structure more
resembles two pairs of arms, or doubled arms. Another fault is
the lack of ``morphological longevity'' in the models of series D: the
spiral structure disperses in just a few bar rotation periods. The
models with a moderate halo have more equally spaced and longer
lasting spiral structure. A common trend in their long-term evolution
is that when the spiral arms weaken owing to gas flow, the minor axis
arms become relatively stronger than the major axis arms. This causes
some models to become essentially two-armed in their later
phases. Both model series produce slightly too pointed and elongated
inner rings. Because the overall fit of the spiral morphology is
better and due to its longevity, we prefer model M6.7 over model D6.7.   

\begin{figure*} 
\resizebox{\hsize}{!}{\includegraphics{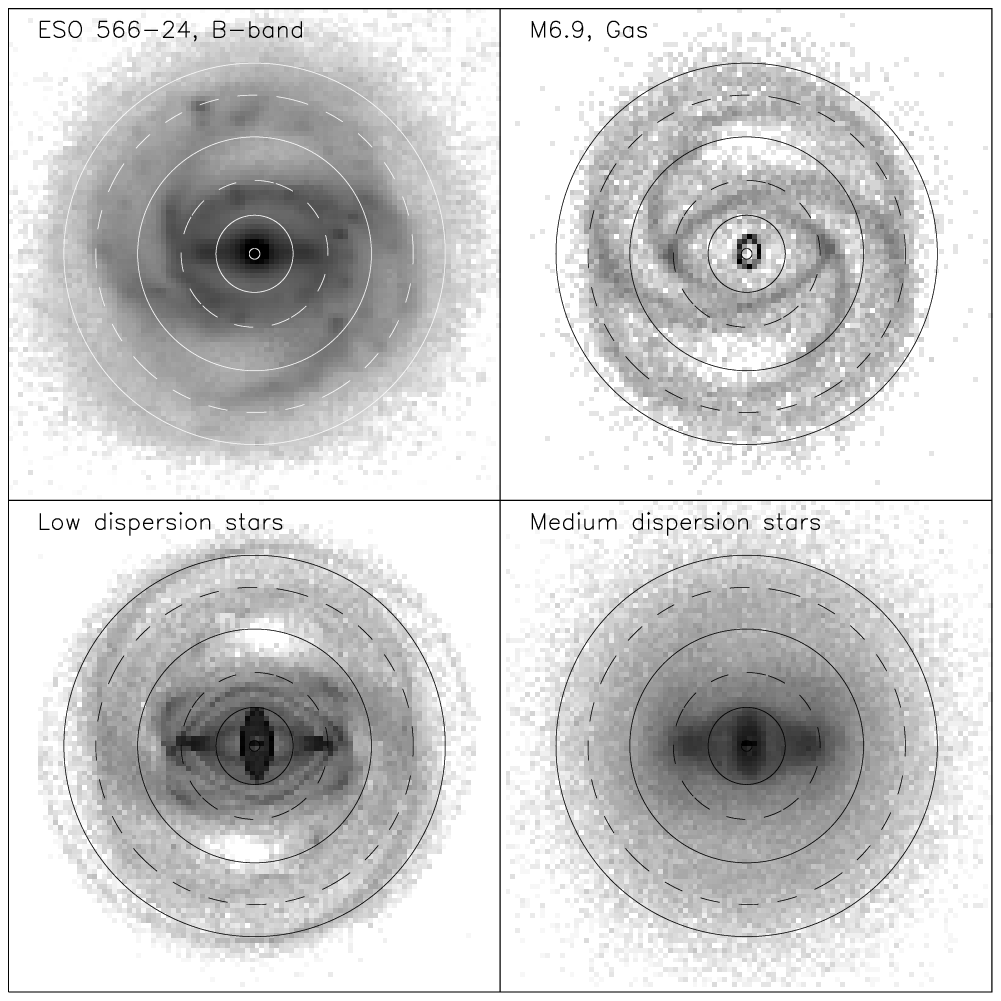}}  \hfill
\caption{The resonance radii vs. morphology. The top left frame shows the
  resonance radii of model M6.9 plotted over the deprojected $B$-band
  image of ESO 566-24, in the top right frame they are plotted over the
  gas morphology of the model. The bottom frames show the morphologies of 
  two populations of non-colliding test particles. In the low dispersion
  case, the velocity dispersion is about 1\% of the local circular
  velocity, while in the medium dispersion case it is about 30\% of the
  circular velocity. The circles drawn with a continuous line
  show the inner Lindblad resonance, the corotation resonance, and the
  outer Lindblad resonance, whereas the circles drawn with a dashed
  line shows the inner and outer 4/1-resonance radii. To increase
  resolution, the simulation images are made by summing particle
  positions of 10 different timesteps after the bar has reached its
  full strength. The width of
  the frames is $120 \arcsec$.}
\label{esoreso} 
\end{figure*}

We have tried to improve the pattern speed estimate by making several
simulations around model M6.7. Judging by the spiral morphology and
the size of the inner ring, we derived $\Omega_b = 6.9 \pm 0.5 \
\mbox{km s}^{-1}\mbox{arcsec}^{-1}$, or about $33.7 \pm 2.3 \ \mbox{km
  s}^{-1}\mbox{kpc}^{-1}$. Fig.~\ref{esoreso} compares this
morphological best-fitting model M6.9 with the $B$-band morphology of
ESO 566-24. The resonance radii corresponding the epicycle
approximation are indicated as circles. The nuclear
ring is between two inner Lindblad resonances ($1.3\arcsec$ and
$9.4\arcsec$) and the inner ring is close to the inner 4/1-resonance
($17.9\arcsec$). The four-armed spiral structure is between the inner and
outer 4/1-resonances ($38.7\arcsec$). The OLR radius ($46.5\arcsec$)
is outside the spiral structure. When compared with the deprojected
image of ESO 566-24, we can see that the bar ends well before the 
corotation resonance radius ($28.5\arcsec$), near the inner 4/1-resonance. Using
the bar radius $18\arcsec \pm 2\arcsec$, this gives $r_{CR}/r_{bar} =
1.58$. Combining the uncertainties in the determination of $r_{bar}$
and $\Omega_b$, we finally get $r_{CR}/r_{bar} = 1.6 \pm 0.3$.  

The morphology of two ``stellar'' populations consisting of
non-colliding test particles is also shown in Fig.~\ref{esoreso}. Both
the low and medium velocity dispersion components (neither of these
components should be expected to represent the self-gravitating
stellar populations but more as a tracer of dynamics) have a strong
contribution of orbits resembling the closed $x_2$-orbits (aligned
perpendicular with respect to the bar) of analytical bar potentials
\citep{contopoulos89b}. Here, we call these orbits $x_2$-like orbits
instead of $x_2$-orbits for two reasons. First, they are orbits of
individual particles in our simulations, not exactly closed orbits
found by iteration. Second, our potential differs from analytical bar
potentials, because it is calculated from NIR-photometry. Thus, our
bar is not straight (this causes the orientation of the inner orbits
to be tilted with respect to the outer parts of the bar) and the
contribution of the spiral arms is also included. Of course, the
situation is somewhat similar in real galaxies, which do not
accommodate idealistic bars. In addition to $x_2$-like orbits, the
presence of $x_1$-like orbits (aligned parallel with respect to the
bar) can be seen in both populations. It is evident that there is no
support for a strong bar structure outside the inner 4/1-resonance. In the
medium dispersion case, there is a low density feature with a
minor-to-major axis ratio of about 0.5, that is aligned with the bar
and reaches almost to corotation. However, it is more like an oval
or a lens than a bar.  

\begin{figure*} 
\resizebox{\hsize}{!}{\includegraphics{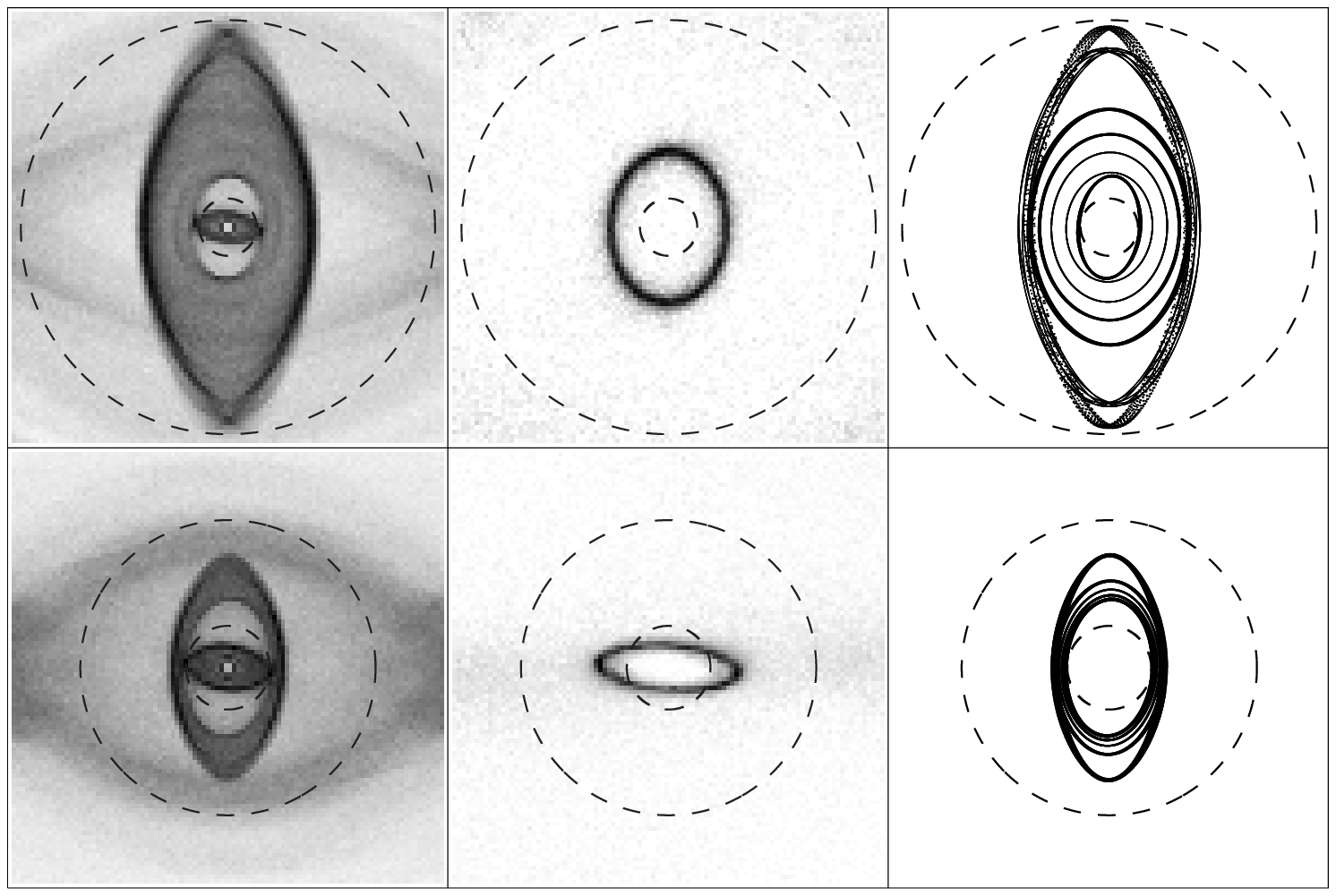}}  \hfill
\caption{Stellar and gaseous nuclear rings vs. $x_2$-like orbits. The
  top row shows model M6.9, the left frame shows the low velocity
  stellar particles, the middle frame the gaseous nuclear ring, and
  the right frame selected stellar particle orbits. The bottom row shows the
  same for model M9.8. Outer and inner ILRs are indicated by circles
  drawn with a dashed line. The width of the frames is $20 \arcsec$.}
\label{ringorbits} 
\end{figure*}

In Fig.~\ref{ringorbits} we compare the
nuclear morphology of low dispersion collisionless populations in two
different bar pattern speeds with the corresponding gas morphology and
we also show some selected orbits of individual stellar particles. In
model M6.9, there is a ``stellar'' nuclear ring, which almost reaches
the outer ILR radius of the epicycle approximation. The gaseous
nuclear ring is considerably smaller. A possible
reason for this difference is the crossing orbits, where colliding
particles cannot stay. The outer
$x_2$-like orbits are crossing both a considerable population of
$x_1$-like orbits and other $x_2$-like orbits. Furthermore, the
outermost orbits do not form (almost) closing loops, although they
retain a perpendicular orientation with respect to the bar. The region
suitable for non-crossing $x_2$-like orbits corresponds quite well with
the size of the gaseous nuclear ring. In model M9.8 with higher pattern 
speed, the $x_2$-like orbits do not reach as close to the linear outer ILR
radius. Their domain is also much narrower, which could explain why
the gaseous nuclear ring has the orientation of $x_1$-like orbits. 

We followed particle orbits in the region of the four-armed spiral and
found a large number of particles librating around Lagrangian points
$L_4$ and $L_5$ \citep{binney87}, which are close to the bar minor
axis. However, the spiral structure cannot be explained by any single
orbit type.  

\begin{figure*} 
\resizebox{\hsize}{!}{\includegraphics{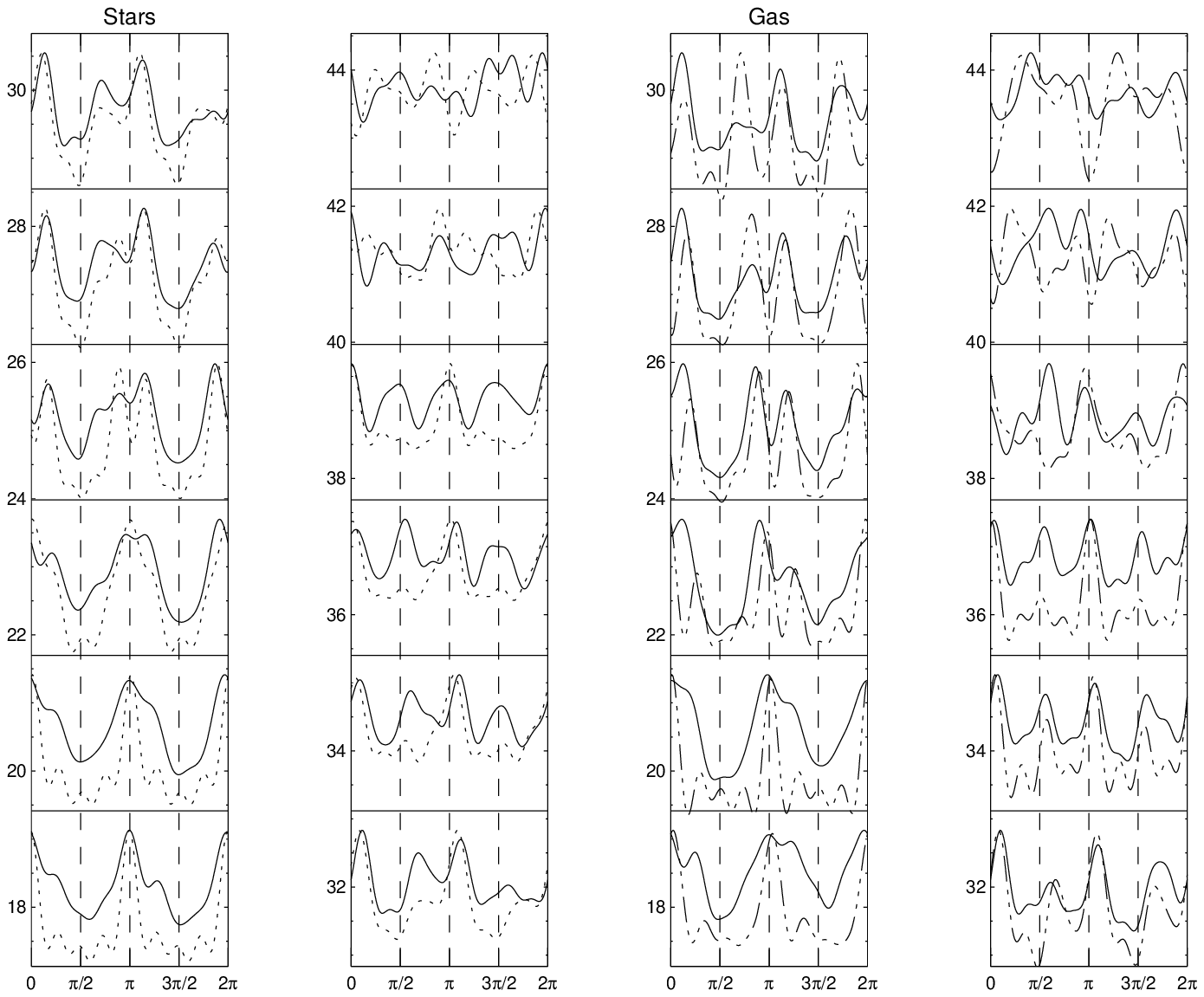}}  \hfill
\caption{The azimuthal variations of surface density at different
  radii (normalized to mean density at this distance) constructed
  from the Fourier decomposition of images. Each curve is calculated
  for the radius indicated by a line below it. The phase angle of the bar
  is zero. The two frames on the left side compare non-colliding test
  particles (low dispersion case) with the $H$-band image, and the frames on
  the right side compare gas particles with the $B$-band image.}
\label{arms} 
\end{figure*}

Fig.~\ref{arms} shows the azimuthal variations of relative surface
density at different radii both in observations and in our best-fitting
model. We can see that although the $m=2$ component is dominating near
the bar end, there are soon signs of an $m=4$
pattern. In the midst of the spiral structure, at about $35 \arcsec$,
all four arms are of about equal strength in the $H$-band image, and the
arm-interarm density contrast can be locally almost as high as 2.0 (about 0.7
magnitudes). Such high values in NIR-images have been found in only
a few galaxies, such as M 51 \citep{rix93}. Comparison between the gas
component of model M6.9 and the $B$-band morphology of ESO 566-24 shows that
the locations of intensity maxima are quite well fitted in the spiral
arms. In the stellar component, the match is worse, which is not
surprising since the model is not self-consistent.  

\subsubsection{Kinematics}
\label{kinematics}

In our previous modelling object, IC 4214, the fine details of the
zero velocity contour of the velocity field were valuable in
determining the bar strength and other model parameters. Even though
the non-axisymmetric perturbation in ESO 566-24 is considerably
stronger than in IC 4214, the zero velocity contour is rather
featureless. This is due to the unfavourable orientation of the bar with
respect to the line of nodes. 

On the other hand, the ``rotation curves'' constructed using the
method introduced by \citet{warner73} shows the effect of non-circular
velocities. The clearest features affected by the non-axisymmetric
perturbation are located inside $10 \arcsec$, just inside the peak of
$Q_T$, the relative non-axisymmetric force. Furthermore, the flatness
of the outer rotation curve clearly indicates that at least some amount
of halo is needed (supposing constant M/L-ratio).

\begin{figure*} 
\resizebox{\hsize}{!}{\includegraphics{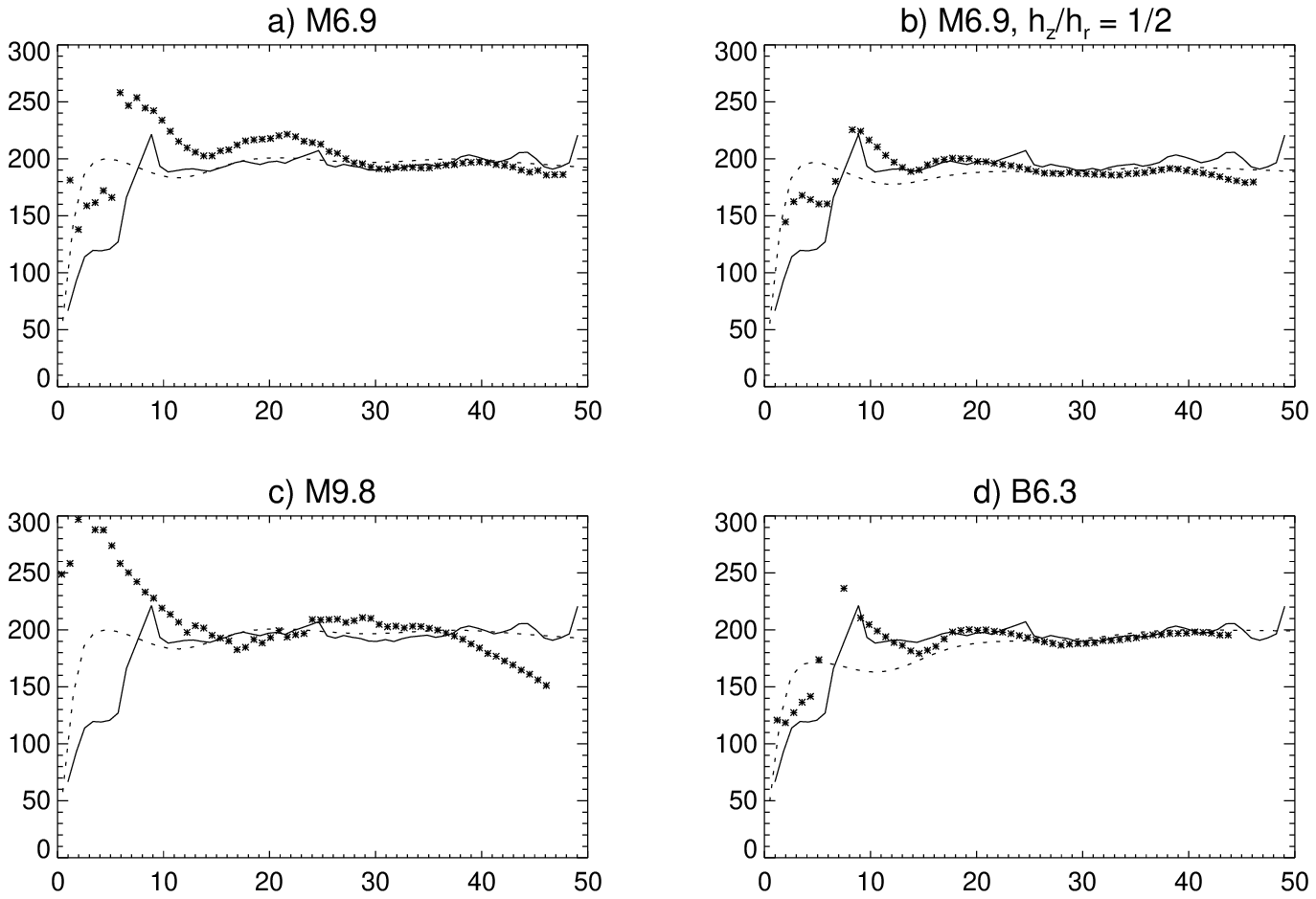}}  \hfill
\caption{Comparison between the observed and modelled kinematics. The
  continuous line shows the observed ``rotation curve'', derived from
  the velocity field using the method by \citet{warner73}. The dotted
  line shows the axisymmetric rotation curve based on the mass model
  and asterisk symbols show the simulated rotation curve constructed
  in the same way as the observed one.}
\label{warvert} 
\end{figure*}

Fig.~\ref{warvert} shows comparison of selected rotation curves constructed
from the modelled velocity field (data from 10 time steps are summed)
with the observed one. The observed velocity field does not have
velocity information from the whole disk area, for example, the
four-armed spiral structure is incompletely covered. On the other
hand, many of our models include gas particles practically in the
whole disk area. This is potentially dangerous when it comes to
comparing the observed and modelled kinematics. To
diminish this problem we adopted a density threshold to the modelled
velocity field. In some cases, this can have a notable effect on the
inner part of the modelled rotation curve. 

Our morphologically fitted ``standard'' model, M6.9 (frame a), gives a
rather good fit with the position of the velocity jump near $10
\arcsec$. The inner parts show also a similar trend as the observed
kinematics. However, the rotation amplitudes do not fit as
well. Whereas the outer rotation amplitude is well fitted, the
modelled velocity is too high inside $25 \arcsec$. A model with a
thicker disk (frame b) gives a better fit, but the rotation amplitude inside
$6\arcsec$ is still too high, by about 40 $\mbox{km \ s}^{-1}$. This
problem cannot be solved by changing the disk thickness or the
pattern speed. A considerable increase of pattern speed destroys the two-stage
structure of the rising part of the rotation curve, as is highlighted
by the kinematics of the model M9.8 (frame c).

To get better kinematical fit, we tried changing the relative masses
of bulge, disk and halo (the radial profiles were not modified). In
model B6.3 (frame d), where $(M/L)_\mathrm{B} = 0.86$,
$(M/L)_\mathrm{D} = 1.02$ and $v_\infty = 219 \ \mbox{km s}^{-1}$ we
could get the difference between the inner velocities as low as $10 -
20 \ \mbox{km s}^{-1}$. With the pattern speed $6.3 \ \mbox{km s}^{-1}
\mbox{arcsec}^{-1}$, which sets the inner 4/1-resonance to
approximately same distance as in model M6.9, the inner and nuclear
rings have about the same size as in model M6.9. However,
the four-armed spiral is not well fitted: instead of long spiral arms,
we get several threaded arms. This is due to the increased halo
contribution. If we further decrease the relative
contribution of the bulge in order to reach the observed rotation
amplitude, the nuclear ring changes its shape and orientation, by becoming
dominated by $x_1$-like orbits. Its morphology is then rather similar to
the nuclear ring in model M9.8, and the central kinematics shows a
peak instead of a drop. Thus, combining the morphological and
kinematical comparison, model M6.9 is our best-fitting solution for
ESO 566-24.  

The behaviour of the determined rotation curve inside $10 \arcsec$ is
related to two factors: the orientation of the bar, i.e.\ its major
axis is close to the kinematical minor axis, and to the orientation of $x_1$-
and $x_2$-like orbits with respect to the bar. In the case of pure
elliptical orbits with major or minor axis parallel with the
kinematical major axis, the resulting rotation curve is either below
(orbit's major axis parallel with the kinematical major axis) or above
(orbit's minor axis parallel with the kinematical major axis) the
local circular velocity \citep{teuben2002}. In barred galaxies, the
shapes of orbits deviate from pure ellipses, but this approximation is
still helpful to understand the general trends in the kinematics of ESO
566-24. Now, the low velocities in the region inside $6\arcsec$ are
caused by $x_2$-like orbits, whose major axis is perpendicular with
respect to the bar major axis. The velocity of these orbits near the
bar minor axis (the major axis of the orbits), is much lower
than the local circular velocity of the azimuthally averaged rotation
curve. This is also close to the kinematical major axis, and thus the
determined rotation curve in this region is below the axisymmetric
rotation curve calculated from the mass distribution. 

Similarly, the velocity
peak near $10 \arcsec$ is caused by $x_1$-like orbits, which are
parallel to the bar. In this distance, they are near their minor axis, and the
velocity is higher than the local circular velocity. The location of
this feature depends on the extent of $x_2$-like gas particle
orbits. This is in part dependent on the pattern speed and the
amplitude of the nonaxisymmetric perturbation. Also evolutionary
effects are present: the jump can move inwards when the size of the
nuclear ring decreases. In some simulations the jump disappears when
this region is depopulated of gas particles. The importance of
$x_2$-like orbits is further demonstrated by the kinematics of models
which lack them: the plateau is replaced by a strong peak (M9.8 for example).

\section{Discussion}
\label{discussion}

\subsection{Spiral morphology in barred galaxies}
\label{m4}

The original form of the
Hubble classification of galaxies \citep{hubble26b} divided spiral
galaxies into subclasses a, b and c, depending not only the relative
size of the bulge, but also on the pitch angle of the spiral arms
and the degree of resolution in the arms. Subsequent modifications or
extensions \citep{devaucouleurs59,sandage61,sandage94,buta95a}, have partially
removed the limitations of the original classification, e.g.\ by providing
description of ring and pseudoring morphologies. A different kind of
spiral classification scheme was developed by
\citet{elmegreen82a,elmegreen87}. Simplifying their classification,
most spiral galaxies can be divided into grand design, multiple-armed, and
flocculent spiral galaxies. Grand design spirals have a global,
usually two-armed spiral structure. In multiple-armed galaxies, the
spiral structure starts as two-armed, but the arms bifurcate later to
several outer arms. Flocculent spirals have fragmented or
chaotic spiral structure. 

A classic case of a barred spiral galaxy is NGC 1300, where a two-armed spiral
structure starts from the ends of the bar. Quite often, the spiral
arms emerge from an inner ring that surrounds the bar. There are also
cases where the spiral arms start offset with respect to the ends of
the bar. This can happen both in the leading and trailing quadrants
of the bar (supposing that the spiral structure is trailing). Two-armed grand
design structure is not the only possibility in barred galaxies.
Different kinds of arm multiplicities also exist: arm bifurcation at
large radius, doubled arms (one or both), and even genuine multiple-arm
structure. Flocculent barred galaxies are rare but known \citep{buta95a}.
According to \citet{elmegreen95a}, the length of the
two-armed phase correlates with bar length: it reaches about twice the
bar radius. 

Most work on galaxy morphology has been done using $B$-band plates. 
A major weakness of this is that the blue wavelength region
is affected by dust extinction and the young stellar population, which 
usually constitutes only a minor part of the mass distribution. On the other
hand, in the NIR most of the light comes from the old stellar
population. It has been shown that bars are more often detected
in the NIR \citep{eskridge2000}, and there are cases where the
multiple-arm or flocculent spiral morphology of visual images
disappears and a two-armed grand design spiral emerges in the NIR
\citep{block91,block96,block99b}. It has sometimes been claimed that
this kind of dualistic morphology is universal
\citep{block99a}. However, \citet{eskridge2002},
who used a large galaxy sample, demonstrated that dramatical
differences between visual and NIR morphology are exceptions, and that in
general galaxy morphology is quite similar in these two wavelength regions. 

The study by \citet{eskridge2002}, based on the Ohio State University
Bright Galaxy Survey (hereafter OSU-survey), includes several barred galaxies
where the multiple-arm structure is seen also in NIR images. The
examples include all variations of arm multiplicities from arm
bifurcations to arm doublings and genuine multiple-armed spirals. For
example, there are three-armed spiral galaxies, e.g.\ NGC 5054
(OSU-survey, another example is NGC 7137
\citealt{grosbol99}). Although there are several four-armed galaxies
in the OSU sample, none of them is similar to ESO 566-24. Most cases
are arm bifurcations at large radius or doubled arms, which are close
to each o her. Even in cases where the extra arms do not form as
bifurcations or doublings of an $m=2$ spiral, they are usually weaker
than the dominating two arms.  

ESO 566-24 with its very regular four-armed spiral structure is really
exceptional. The four arms have roughly the same length, so they do
not appear to have formed as a bifurcation of a usual two-armed structure. 
Furthermore, they are
so far from each other (at $35 \arcsec$, the phase difference
between the arms is about $90\degr$), that arm doubling is clearly not the
case. To crown the status of ESO 566-24 as a genuine $m=4$ spiral, the
arms have comparable strength also in the NIR. Thus, one
cannot choose any pair of arms, say those starting near the bar ends,
as two major arms. In the OSU-survey, the galaxy with the closest
resemblance to ESO 566-24 is NGC 613. It has altogether five spiral
arms, but the ``extra'' arms are not as strong and regular as in ESO 566-24.  

Some barred galaxies have features
called plumes, i.e.\ short arcs besides the main spiral arms in the leading
side with respect to the bar \citep{buta84}. The best-known
case is NGC 1433 \citep{buta86b,ryder96}. It is possible that plumes
could be related to four-armed spiral structure, for example, they could be
remnants of additional spiral arms. However, such features can form in
gasdynamical simulations as result of doubled spiral arms
\citep{byrd94}. It would be interesting to thoroughly compare ESO
566-24 with NGC 1433.

Although our best-fitting models can reproduce the four-armed spiral
rather well, and give constraints to essential model parameters, they
do not directly unveil the origin of such structure. Instead, even in
the best case, our models are snapshots of the current stage of the
evolution of ESO 566-24. It is possible that the location of
resonances can give a clue of the origin of the four-armed
spiral. Based on our models, it seems that four-armed structure is
approximately confined between inner and outer 4/1-resonances. Thus
the situation could be analogous to two-armed spirals, which are
sometimes suggested to end near OLR
\citep[e.g.][]{elmegreen89b}. Note however, that at least for non-barred
galaxies, other solutions, which put the end of the spiral structure
inside corotation or even the inner 4/1-resonance, have also been suggested
\citep{patsis94,patsis97a}.   

\subsection{Rings}

A connection between rings and resonances was perhaps
first suggested by \citet{schommer76}, who identified two rings in NGC
4736 with inner and outer Lindblad resonances. Later, sticky particle
simulations by \citet{schwarz81,schwarz84a,schwarz84b} showed that gas can
accumulate near the resonance radii. These and other studies
\citep{combes85,byrd94,rautiainen2000} have identified nuclear rings
with orbits near the ILR, inner rings near the  inner
4/1-resonance and outer rings near the OLR. Two different observed
populations of outer rings and pseudorings \citep{buta95a}
are nicely explained by orbits inside and
outside OLR: inside OLR the major axes of the orbits belonging to the
dominating orbit family are oriented perpendicular 
to the bar, whereas outside OLR they are oriented parallel to the bar.

In many studies, the nuclear rings in
single bar potentials have been identified with $x_2$-orbits, which
are perpendicular to the bar. When the non-axisymmetric perturbation
is weak, and the deviations from circular motion are small, the
outermost stable $x_2$-orbits reach the ILR (or the outer ILR when there are
two inner Lindblad resonances) \citep{contopoulos89b}. However, when
the strength of the perturbation is increased, the linear
approximation is not valid, and the outermost extent of
$x_2$-orbits moves inwards. In fact, some authors have redefined the
ILR-radius so that it means the radius with the outermost stable $x_2$-orbits
\citep{vanalbada82,athanassoula92a,miwa98}. 

Recently, \citet{regan2003} have studied the formation and evolution of
nuclear rings using a hydrodynamics code with an analytic bar
potential. Their conclusion is that the nuclear rings are not related
to inner Lindblad resonances, but instead they form as an interaction
between gas streaming along $x_2$- and $x_1$-like streamlines. Furthermore,
the nuclear rings in their simulations did not have a steady size but
shrank throughout the simulation. Such behaviour can be seen also in
our simulations, but the change in the size of the nuclear ring
becomes very slow in the later phases of the simulations. This can be
due to lack of interaction with particles on $x_1$-like orbits, which
have become depopulated.

We have tested the hypothesis by Regan and Teuben that the nuclear
rings form as interaction of two different gas streamlines by making
simulations where all the particles initially resided in the domain of
$x_2$-like orbits. A clear nuclear ring formed without interaction of two
streamlines. In fact, a nuclear ring may form even without any gas: the
redistribution of non-colliding test particles by a rotating bar also forms
a ``concentration'' of particles near the outermost $x_2$-like orbits. 
Rings can form also in self-gravitating stellar particles of N-body
simulations \citep{rautiainen2000,athanassoula2002a}. In ESO 566-24,
the inner ring is clearly present in the $H$-band, and
corresponding examples exist also for outer and nuclear
rings. \citet{byrd94} speculated that stellar rings could be remnants
of star formation in gaseous rings, but it is also possible that a
ring forms as a response of the old stellar population to a
non-axisymmetric potential.

If the rings are connected to orbits in the presence of a
non-axisymmetric bar potential, then the shapes of the rings should
correlate with the bar strength. \citet{buta2002} compared the shapes
of the inner rings in a few galaxies with the $Q_b$-parameter, and did not
find a clear correlation. Galaxies with roughly similar $Q_b$ values
can have inner rings of different shapes, ranging from virtually
circular to highly elongated. This can be explained in 
several ways. First, $Q_b$ measures the peak strength of the bar,
which often takes place well inside the inner ring, i.e.\ $Q_b$ may
not be representative of tangential force in the ring region. This
effect can be seen in action in Figs. 5 -- 7: in the higher pattern
speed models, the inner ring is located closer to centre, in a region with
higher $Q_T$, which explains the more elongated shape ($Q_b$ is
constant in each pattern speed series). Second, the
$Q_b$ values were calculated from NIR photometry, and thus the
possible contribution of dark matter was omitted. 

\subsection{Pattern speed dilemma}
\label{patternspeed}

Our best-fitting models have pattern speeds which place the corotation
resonance at about 1.6 times the bar radius. This is considerably
larger than the often cited value $r_{CR}/r_{bar} = 1.2 \pm 0.2$. For
IC 4214 we found a value of $1.4 \pm 0.2$ \citep{salo99}, thus just barely
fitting within the fast bar regime although it has an early-type
morphology. This raises a question if our modelling method is biased so
that it gives too high values. Our preliminary results with NGC 4314
suggest otherwise; for this galaxy we get $r_{CR}/r_{bar}$ close to one.

We extended a previous study by \citet{elmegreen96d} by a brief search
of the literature. We found pattern speed determinations of individual
barred galaxies giving values ranging from below 1 to over 2, the
average value being about 1.35 (median value was 1.3). The values based on
the Tremaine-Weinberg method tend to be close to 1, although
\citet{aguerri2003} found considerably higher values for galaxies NGC
1440 ($1.6^{+0.5}_{-0.3}$) and NGC 3412
($1.5^{+0.6}_{-0.3}$). Furthermore, the Tremaine-Weinberg method has
been used almost exclusively with SB0-galaxies. ESO 566-24 is of Hubble
type SB(r)b, later than most galaxies whose pattern speed has been
determined with the Tremaine-Weinberg method. \citet{elmegreen85} have
suggested that early-type galaxies have longer bars than late-type
galaxies in both the physical size and with respect to resonance
locations. Thus, it seems possible that the high value found in our
study could somehow be related to Hubble type.

A point in favour of higher pattern speeds is the existence of dust
lanes inside the bar. According to \citet{athanassoula92b} the offset
dust lanes form when the galaxy has an ILR and their shape corresponds to the
observations when the corotation radius is $1.2 \pm 0.2$
times the bar radius. However, the dust lane argument for fast bars may
not be as strong as it seems: \citet{lindblad96b} found two possible
pattern speeds for NGC 1300, giving $r_{CR}/r_{bar} = 1.3$ and
$r_{CR}/r_{bar} = 2.4$, both producing good dust lane
morphology. 

N-body studies have shown that pattern speeds of the bar
and spiral arms can be different, structures being either separate or in
a non-linear mode coupling
\citep{sellwood88,masset97,rautiainen99}. \citet{rautiainen2000}
demonstrated that rings can form also in self-gravitating simulations
with two or more modes with different pattern speeds. Sometimes, the
presence of two strong modes caused cyclic evolution in the shape and
orientation of the ring. At least in the case of nuclear rings, this
can be attributed to orbit loops \citep{maciejewski2000} in a
potential with two rotating components \citep{rautiainen2002}. So, in
principle it is it possible that ESO 566-24 could be modelled with two
modes: the inner region having a higher pattern speed and the four-armed
spiral a pattern speed in the range of our single pattern
models. However, in this case the low value for the pattern speed is
based not only on the modelling of the four-armed spiral structure,
but also the central morphology is better modelled with low pattern speed:
the inner and nuclear rings have approximately the right size. When
the pattern speed is higher, both become too elongated or
disappear altogether.     

\subsection{Halo contribution}
\label{halocontrib}

Early N-body simulations of barred galaxies \citep{sellwood81}
exhibited deceleration of the bar rotation due to interaction with
the outer disk. This made the corotation radius to move outwards, but
it was compensated by the growth of the bar, i.e.\ the bar
always ended close to corotation. However, interaction between bar and
dark halo can introduce bar slowdown that cannot be compensated by
bar growth.  \citet{weinberg85} studied the dynamical friction
between a bar and halo analytically and also by a ``semi-resticted''
N-body simulation (particles are not self-gravitating), and found
that it causes the bar to lose most of its angular
momentum in just a few bar rotations. \citet{little91a} studied bar
slowdown by self-gravitating N-body models, where both the disk and
halo were two-dimensional. The bar deceleration in their model was
considerably lower, only by a factor of two during 10 Gyr, half of
which was due to interaction with the outer disk. On the
other hand, \citet{hernquist92} who used an analytical bar with a
self-gravitating halo, found a slowdown rate corresponding to
that found by Weinberg.  

At least seemingly confusing results have been
obtained also with fully self-consistent N-body models of bar--halo
interaction. \citet{debattista98,debattista2000} find that in models
with centrally concentrated haloes, the bar slows down dramatically
($r_{CR}/r_{bar} > 1.4$), unless the halo has unrealistically high
angular momentum in the same direction as the disk.
\citet{valenzuela2002} made simulations using adaptive grid refinement
(the potential grid was subdivided in high density regions), which
increased the gravity resolution. They found that the bar
pattern speed was almost constant and that the stellar disk lost only
5 -- 10\% of its angular momentum to the halo. Furthermore,
they could reproduce the dramatic bar slowdown when the grid
resolution was decreased. Recently, \citet{athanassoula2003} found
that when the halo velocity dispersion is increased, the bar slowdown
rate also decreases. 

The preceding discussion indicates that it is
questionable whether bar pattern speed can be used as an argument for
high or low halo-to-disk mass ratios.
If we compare our best-fitting mass model with the simulations of
\citet{debattista98} with a non-rotating halo, and adopt their parameter
$\eta=(v_{disk}/v_{halo})^2$ at the maximum of disk rotation curve, we
find $\eta=3.65$. In their Fig. 2, this would correspond to
$r_{CR}/r_{bar} = 1.3 - 1.4$ at the final equilibrium state. On the
other hand, if we include bulge to $v_{halo}$, we get $\eta=1.39$,
corresponding $r_{CR}/r_{bar} >2$ in
\citet{debattista98}. \citet{debattista2000} found that a halo which
rotates in the same sense as the bar can induce lower bar slowdown
rate than non-rotating or retrograde halos. When taking into account
that bulges rotate, our result $r_{CR}/r_{bar} \approx 1.6$ is
not in disagreement with results of \citet{debattista98,debattista2000}. 
However, the situation is the same if we compare with
simulations by \citet{valenzuela2002}: in their models the corotation
resonance radius is 1.2 to 1.7 times the bar radius, and our value is
near the upper limit of their range. 

In our best fitting model (M6.9), the dark halo contributes less than
20\% of the mass inside $r_{2.2}$. The rotation curve is dominated by
luminous matter even inside the whole disk region. Thus, our results with
ESO 566-24 are in accordance with estimates of disk
mass contribution by dynamical modelling
\citep[e.g.][]{weiner2001b}, but disagree with cold dark matter
(CDM) cosmological N-body simulations, which produce cuspy
halos \citep[e.g.][]{navarro96}%,navarro2000b,navarro2000a}.   

\section{Conclusions}
\label{conclusions}

We have constructed dynamical models for the four-armed barred spiral
galaxy ESO 566-24. The mass distribution of the different components
is based on near-IR photometry with the exception of a possible dark
halo, whose amount was chosen to fit the rotation curve at large radii.
We have been able to construct a model which reproduces the
observed four-armed morphology. Also the main kinematic
characteristics are seen in our models. The main conclusions are as
follows. 

1. Two major factors affecting the simulated morphology and kinematics
   are the strength of the non-axisymmetric perturbation and the
   pattern speed of the bar. The former depends on the mass and the
   thickness of the disk.

2. Both the kinematical and morphological fits are better when a
   moderate halo component is included. On the other hand, a
   dominating halo can be ruled out, because then the gas response to
   the disk gravitational potential is too weak. Thus, the
   contribution of the luminous matter dominates the rotation curve in
   the whole optical disk.  

3. The four-armed spiral could be produced between the
   inner and outer 4/1-resonances. The inner ring is close to the inner
   4/1-resonance and the nuclear ring is between two inner Lindblad
   resonances. The size
   of the nuclear ring seems to correlate with the size of the region
   suitable for non-crossing $x_2$-like orbits.

4. If the bar has the same pattern speed as the four-armed spiral,
   then it rotates rather slowly; the corotation radius is about
   $1.6 \pm 0.3$ times the bar radius (including uncertainties both in
   the bar radius and pattern speed), which is higher
   than the often cited value of $1.2 \pm 0.2$. The existence of several
   other determined high values for this ratio suggests that
   there is more scatter in the $r_{CR}/r_{bar}$ ratio than is
   sometimes stated. It should also be noted that most bar
   pattern speed determinations have been made for galaxies of earlier
   Hubble type than ESO 566-24.

5. The rather high value of $r_{CR}/r_{bar}$ is difficult to be
   disputed by pattern speed multiplicity. In addition to the
   spiral structure, the nuclear and inner rings can be reproduced
   in the right scale with a single pattern speed. Adopting
   considerably higher pattern speeds, which would be consistent with
   the bar ending near corotation, produce significantly worse-fitting
   ring morphology. Also, the kinematics of the central parts deviate
   more from the observed velocities. 

6. The relatively slow bar rotation rate can be due to interaction
   between the bar and the spheroidal component (halo and bulge). The
   exceptional $m=4$ spiral morphology can be related to the low bar
   rotation rate and the importance of the spheroidal component.

7. The curious shape of the inner rotation curve can be explained by
   combined effects of $x_1$- and $x_2$-like orbits. 

\section*{Acknowledgments}

This work has been supported by V\"ais\"al\"a Foundation
and by the Academy of Finland. RB acknowledges the support of
NSF grant AST-0205143 to the University of Alabama.
This work made use of data from the Ohio State University Bright
Spiral Galaxy Survey, which was funded by grants AST-9217716 and
AST-9617006 from the United States National Science Foundation, with
additional support from the Ohio State University.  

\bibliographystyle{mn2e}
\bibliography{astrobib}

\begin{thebibliography}{}

\bibitem[\protect\citeauthoryear{{Aguerri}, {Debattista} \&
  {Corsini}}{{Aguerri} et~al.}{2003}]{aguerri2003}
{Aguerri} J. A.~L.,  {Debattista} V.~P.,    {Corsini} E.~M.,  2003, MNRAS, 338,
  465

\bibitem[\protect\citeauthoryear{{Athanassoula}}{{Athanassoula}}{1992a}]{athan%
assoula92a}
{Athanassoula} E.,  1992a, MNRAS, 259, 328

\bibitem[\protect\citeauthoryear{{Athanassoula}}{{Athanassoula}}{1992b}]{athan%
assoula92b}
{Athanassoula} E.,  1992b, MNRAS, 259, 345

\bibitem[\protect\citeauthoryear{{Athanassoula}}{{Athanassoula}}{2002}]{athana%
ssoula2002b}
{Athanassoula} E.,  2002, ApJL, 569, L83

\bibitem[\protect\citeauthoryear{{Athanassoula}}{{Athanassoula}}{2003}]{athana%
ssoula2003}
{Athanassoula} E.,  2003, MNRAS, 341, 1179

\bibitem[\protect\citeauthoryear{{Athanassoula} \& {Misiriotis}}{{Athanassoula}
  \& {Misiriotis}}{2002}]{athanassoula2002a}
{Athanassoula} E.,  {Misiriotis} A.,  2002, MNRAS, 330, 35

\bibitem[\protect\citeauthoryear{{Ball}}{{Ball}}{1992}]{ball92}
{Ball} R.,  1992, ApJ, 395, 418

\bibitem[\protect\citeauthoryear{{Binney} \& {Tremaine}}{{Binney} \&
  {Tremaine}}{1987}]{binney87}
{Binney} J.,  {Tremaine} S.,  1987, Galactic dynamics.
Princeton, NJ, Princeton University Press, 1987

\bibitem[\protect\citeauthoryear{{Block}, {Elmegreen} \& {Wainscoat}}{{Block}
  et~al.}{1996}]{block96}
{Block} D.~L.,  {Elmegreen} B.~G.,    {Wainscoat} R.~J.,  1996, Nature, 381,
  674

\bibitem[\protect\citeauthoryear{{Block} \& {Puerari}}{{Block} \&
  {Puerari}}{1999}]{block99a}
{Block} D.~L.,  {Puerari} I.~.,  1999, A\&A, 342, 627

\bibitem[\protect\citeauthoryear{{Block}, {Puerari}, {Frogel}, {Eskridge},
  {Stockton} \& {Fuchs}}{{Block} et~al.}{1999}]{block99b}
{Block} D.~L.,  {Puerari} I.~.,  {Frogel} J.~A.,  {Eskridge} P.~B.,  {Stockton}
  A.,    {Fuchs} B.,  1999, AP\&SS, 269, 5

\bibitem[\protect\citeauthoryear{{Block} \& {Wainscoat}}{{Block} \&
  {Wainscoat}}{1991}]{block91}
{Block} D.~L.,  {Wainscoat} R.~J.,  1991, Nature, 353, 48

\bibitem[\protect\citeauthoryear{{Buta}}{{Buta}}{1984}]{buta84}
{Buta} R.,  1984, PASAu, 5, 472

\bibitem[\protect\citeauthoryear{{Buta}}{{Buta}}{1986}]{buta86b}
{Buta} R.,  1986, ApJS, 61, 631

\bibitem[\protect\citeauthoryear{{Buta}}{{Buta}}{1988}]{buta88}
{Buta} R.,  1988, ApJS, 66, 233

\bibitem[\protect\citeauthoryear{{Buta}}{{Buta}}{1995}]{buta95a}
{Buta} R.,  1995, ApJS, 96, 39

\bibitem[\protect\citeauthoryear{{Buta}}{{Buta}}{2002}]{buta2002}
{Buta} R.,  2002, ASP Conf. Ser. 275: Disks of Galaxies: Kinematics, Dynamics
  and Peturbations, p.~185

\bibitem[\protect\citeauthoryear{{Buta}, {Alpert}, {Cobb}, {Crocker} \&
  {Purcell}}{{Buta} et~al.}{1998}]{buta98b}
{Buta} R.,  {Alpert} A.~J.,  {Cobb} M.~L.,  {Crocker} D.~A.,    {Purcell}
  G.~B.,  1998, AJ, 116, 1142

\bibitem[\protect\citeauthoryear{{Buta} \& {Block}}{{Buta} \&
  {Block}}{2001}]{buta2001b}
{Buta} R.,  {Block} D.~L.,  2001, ApJ, 550, 243

\bibitem[\protect\citeauthoryear{{Buta} \& {Combes}}{{Buta} \&
  {Combes}}{1996}]{buta96b}
{Buta} R.,  {Combes} F.,  1996, Fundamentals of Cosmic Physics, 17, 95

\bibitem[\protect\citeauthoryear{{Buta} \& {Combes}}{{Buta} \&
  {Combes}}{2000}]{buta2000a}
{Buta} R.,  {Combes} F.,  2000, ASP Conf. Ser. 197: Dynamics of Galaxies: from
  the Early Universe to the Present, p.~11

\bibitem[\protect\citeauthoryear{{Buta} \& {Crocker}}{{Buta} \&
  {Crocker}}{1991}]{buta91b}
{Buta} R.,  {Crocker} D.~A.,  1991, AJ, 102, 1715

\bibitem[\protect\citeauthoryear{{Buta} \& {Crocker}}{{Buta} \&
  {Crocker}}{1993}]{buta93a}
{Buta} R.,  {Crocker} D.~A.,  1993, AJ, 105, 1344

\bibitem[\protect\citeauthoryear{{Buta} \& {Purcell}}{{Buta} \&
  {Purcell}}{1998}]{buta98a}
{Buta} R.,  {Purcell} G.~B.,  1998, AJ, 115, 484

\bibitem[\protect\citeauthoryear{{Buta}, {Purcell}, {Cobb}, {Crocker},
  {Rautiainen} \& {Salo}}{{Buta} et~al.}{1999}]{buta99b}
{Buta} R.,  {Purcell} G.~B.,  {Cobb} M.~L.,  {Crocker} D.~A.,  {Rautiainen} P.,
     {Salo} H.,  1999, AJ, 117, 778

\bibitem[\protect\citeauthoryear{{Buta}, {Ryder}, {Madsen}, {Wesson}, {Crocker}
  \& {Combes}}{{Buta} et~al.}{2001}]{buta2001a}
{Buta} R.,  {Ryder} S.~D.,  {Madsen} G.~J.,  {Wesson} K.,  {Crocker} D.~A.,
  {Combes} F.,  2001, AJ, 121, 225

\bibitem[\protect\citeauthoryear{{Byrd}, {Rautiainen}, {Salo}, {Buta} \&
  {Crocher}}{{Byrd} et~al.}{1994}]{byrd94}
{Byrd} G.,  {Rautiainen} P.,  {Salo} H.,  {Buta} R.,    {Crocher} D.~A.,  1994,
  AJ, 108, 476

\bibitem[\protect\citeauthoryear{{Canzian}}{{Canzian}}{1993}]{canzian93}
{Canzian} B.,  1993, ApJ, 414, 487

\bibitem[\protect\citeauthoryear{{Canzian} \& {Allen}}{{Canzian} \&
  {Allen}}{1997}]{canzian97}
{Canzian} B.,  {Allen} R.~J.,  1997, ApJ, 479, 723

\bibitem[\protect\citeauthoryear{{Combes} \& {Elmegreen}}{{Combes} \&
  {Elmegreen}}{1993}]{combes93}
{Combes} F.,  {Elmegreen} B.~G.,  1993, A\&A, 271, 391

\bibitem[\protect\citeauthoryear{{Combes} \& {Gerin}}{{Combes} \&
  {Gerin}}{1985}]{combes85}
{Combes} F.,  {Gerin} M.,  1985, A\&A, 150, 327

\bibitem[\protect\citeauthoryear{{Combes} \& {Sanders}}{{Combes} \&
  {Sanders}}{1981}]{combes81}
{Combes} F.,  {Sanders} R.~H.,  1981, A\&A, 96, 164

\bibitem[\protect\citeauthoryear{{Contopoulos} \& {Grosb{\o}l}}{{Contopoulos}
  \& {Grosb{\o}l}}{1989}]{contopoulos89b}
{Contopoulos} G.,  {Grosb{\o}l} P.,  1989, A\&AR, 1, 261

\bibitem[\protect\citeauthoryear{{de Grijs}}{{de Grijs}}{1998}]{degrijs98}
{de Grijs} R.,  1998, MNRAS, 299, 595

\bibitem[\protect\citeauthoryear{{de Grijs}, {Peletier} \& {van der Kruit}}{{de
  Grijs} et~al.}{1997}]{degrijs97}
{de Grijs} R.,  {Peletier} R.~F.,    {van der Kruit} P.~C.,  1997, A\&A, 327,
  966

\bibitem[\protect\citeauthoryear{{de Vaucouleurs}}{{de
  Vaucouleurs}}{1959}]{devaucouleurs59}
{de Vaucouleurs} G.,  1959, Handbuch der Physik, 53, 275

\bibitem[\protect\citeauthoryear{{Debattista} \& {Sellwood}}{{Debattista} \&
  {Sellwood}}{1998}]{debattista98}
{Debattista} V.~P.,  {Sellwood} J.~A.,  1998, ApJ, 493, L5

\bibitem[\protect\citeauthoryear{{Debattista} \& {Sellwood}}{{Debattista} \&
  {Sellwood}}{2000}]{debattista2000}
{Debattista} V.~P.,  {Sellwood} J.~A.,  2000, ApJ, 543, 704

\bibitem[\protect\citeauthoryear{{Duval} \& {Athanassoula}}{{Duval} \&
  {Athanassoula}}{1983}]{duval83}
{Duval} M.~F.,  {Athanassoula} E.,  1983, A\&A, 121, 297

\bibitem[\protect\citeauthoryear{{Elmegreen}}{{Elmegreen}}{1996}]{elmegreen96d}
{Elmegreen} B.,  1996, ASP Conf. Ser. 91: IAU Colloq. 157: Barred Galaxies,
  p.~197

\bibitem[\protect\citeauthoryear{{Elmegreen} \& {Elmegreen}}{{Elmegreen} \&
  {Elmegreen}}{1985}]{elmegreen85}
{Elmegreen} B.~G.,  {Elmegreen} D.~M.,  1985, ApJ, 288, 438

\bibitem[\protect\citeauthoryear{{Elmegreen}, {Seiden} \&
  {Elmegreen}}{{Elmegreen} et~al.}{1989}]{elmegreen89b}
{Elmegreen} B.~G.,  {Seiden} P.~E.,    {Elmegreen} D.~M.,  1989, ApJ, 343, 602

\bibitem[\protect\citeauthoryear{{Elmegreen} \& {Elmegreen}}{{Elmegreen} \&
  {Elmegreen}}{1982}]{elmegreen82a}
{Elmegreen} D.~M.,  {Elmegreen} B.~G.,  1982, MNRAS, 201, 1021

\bibitem[\protect\citeauthoryear{{Elmegreen} \& {Elmegreen}}{{Elmegreen} \&
  {Elmegreen}}{1987}]{elmegreen87}
{Elmegreen} D.~M.,  {Elmegreen} B.~G.,  1987, ApJ, 314, 3

\bibitem[\protect\citeauthoryear{{Elmegreen} \& {Elmegreen}}{{Elmegreen} \&
  {Elmegreen}}{1995}]{elmegreen95a}
{Elmegreen} D.~M.,  {Elmegreen} B.~G.,  1995, ApJ, 445, 591

\bibitem[\protect\citeauthoryear{{England}}{{England}}{1989}]{england89}
{England} M.~N.,  1989, ApJ, 344, 669

\bibitem[\protect\citeauthoryear{{Eskridge}, {Frogel}, {Pogge}, {Quillen},
  {Berlind}, {Davies}, {DePoy}, {Gilbert}, {Houdashelt}, {Kuchinski}, {Ram{\'
  i}rez}, {Sellgren}, {Stutz}, {Terndrup} \& {Tiede}}{{Eskridge}
  et~al.}{2002}]{eskridge2002}
{Eskridge} P.~B.,  {Frogel} J.~A.,  {Pogge} R.~W.,  {Quillen} A.~C.,  {Berlind}
  A.~A.,  {Davies} R.~L.,  {DePoy} D.~L.,  {Gilbert} K.~M.,  {Houdashelt}
  M.~L.,  {Kuchinski} L.~E.,  {Ram{\' i}rez} S.~V.,  {Sellgren} K.,  {Stutz}
  A.,  {Terndrup} D.~M.,    {Tiede} G.~P.,  2002, ApJS, 143, 73

\bibitem[\protect\citeauthoryear{{Eskridge}, {Frogel}, {Pogge}, {Quillen},
  {Davies}, {DePoy}, {Houdashelt}, {Kuchinski}, {Ram{\'i}rez}, {Sellgren},
  {Terndrup} \& {Tiede}}{{Eskridge} et~al.}{2000}]{eskridge2000}
{Eskridge} P.~B.,  {Frogel} J.~A.,  {Pogge} R.~W.,  {Quillen} A.~C.,  {Davies}
  R.~L.,  {DePoy} D.~L.,  {Houdashelt} M.~L.,  {Kuchinski} L.~E.,
  {Ram{\'i}rez} S.~V.,  {Sellgren} K.,  {Terndrup} D.~M.,    {Tiede} G.~P.,
  2000, AJ, 119, 536

\bibitem[\protect\citeauthoryear{{Grosb{\o}l} \& {Patsis}}{{Grosb{\o}l} \&
  {Patsis}}{1999}]{grosbol99}
{Grosb{\o}l} P.~J.,  {Patsis} P.~A.,  1999, ASP Conf. Ser. 182: Galaxy Dynamics
  - A Rutgers Symposium, p.~217

\bibitem[\protect\citeauthoryear{{Hernquist} \& {Weinberg}}{{Hernquist} \&
  {Weinberg}}{1992}]{hernquist92}
{Hernquist} L.,  {Weinberg} M.~D.,  1992, ApJ, 400, 80

\bibitem[\protect\citeauthoryear{{Hubble}}{{Hubble}}{1926}]{hubble26b}
{Hubble} E.~P.,  1926, ApJ, 64, 321

\bibitem[\protect\citeauthoryear{{Kent}}{{Kent}}{1986}]{kent86}
{Kent} S.~M.,  1986, AJ, 91, 1301

\bibitem[\protect\citeauthoryear{{Kent}}{{Kent}}{1987}]{kent87}
{Kent} S.~M.,  1987, AJ, 93, 1062

\bibitem[\protect\citeauthoryear{{Knapen}, {Shlosman} \& {Peletier}}{{Knapen}
  et~al.}{2000}]{knapen2000}
{Knapen} J.~H.,  {Shlosman} I.,    {Peletier} R.~F.,  2000, ApJ, 529, 93

\bibitem[\protect\citeauthoryear{{Kranz}, {Slyz} \& {Rix}}{{Kranz}
  et~al.}{2001}]{kranz2001}
{Kranz} T.,  {Slyz} A.,    {Rix} H.,  2001, ApJ, 562, 164

\bibitem[\protect\citeauthoryear{{Laurikainen} \& {Salo}}{{Laurikainen} \&
  {Salo}}{2002}]{laurikainen2002b}
{Laurikainen} E.,  {Salo} H.,  2002, MNRAS, 337, 1118

\bibitem[\protect\citeauthoryear{{Laurikainen}, {Salo} \&
  {Rautiainen}}{{Laurikainen} et~al.}{2002}]{laurikainen2002a}
{Laurikainen} E.,  {Salo} H.,    {Rautiainen} P.,  2002, MNRAS, 331, 880

\bibitem[\protect\citeauthoryear{{Lindblad} \& {Kristen}}{{Lindblad} \&
  {Kristen}}{1996}]{lindblad96b}
{Lindblad} P. A.~B.,  {Kristen} H.,  1996, A\&A, 313, 733

\bibitem[\protect\citeauthoryear{{Lindblad}, {Lindblad} \&
  {Athanassoula}}{{Lindblad} et~al.}{1996}]{lindblad96a}
{Lindblad} P. A.~B.,  {Lindblad} P.~O.,    {Athanassoula} E.,  1996, A\&A, 313,
  65

\bibitem[\protect\citeauthoryear{{Little} \& {Carlberg}}{{Little} \&
  {Carlberg}}{1991}]{little91a}
{Little} B.,  {Carlberg} R.~G.,  1991, MNRAS, 250, 161

\bibitem[\protect\citeauthoryear{{Maciejewski} \& {Sparke}}{{Maciejewski} \&
  {Sparke}}{2000}]{maciejewski2000}
{Maciejewski} W.,  {Sparke} L.~S.,  2000, MNRAS, 313, 745

\bibitem[\protect\citeauthoryear{{Maoz}, {Barth}, {Sternberg}, {Filippenko},
  {Ho}, {Macchetto}, {Rix} \& {Schneider}}{{Maoz} et~al.}{1996}]{maoz96}
{Maoz} D.,  {Barth} A.~J.,  {Sternberg} A.,  {Filippenko} A.~V.,  {Ho} L.~C.,
  {Macchetto} F.~D.,  {Rix} H.~.,    {Schneider} D.~P.,  1996, AJ, 111, 2248

\bibitem[\protect\citeauthoryear{{Masset} \& {Tagger}}{{Masset} \&
  {Tagger}}{1997}]{masset97}
{Masset} F.,  {Tagger} M.,  1997, A\&A, 322, 442

\bibitem[\protect\citeauthoryear{{Merrifield} \& {Kuijken}}{{Merrifield} \&
  {Kuijken}}{1995}]{merrifield95}
{Merrifield} M.~R.,  {Kuijken} K.,  1995, MNRAS, 274, 933

\bibitem[\protect\citeauthoryear{{Miwa} \& {Noguchi}}{{Miwa} \&
  {Noguchi}}{1998}]{miwa98}
{Miwa} T.,  {Noguchi} M.,  1998, ApJ, 499, 149

\bibitem[\protect\citeauthoryear{{Navarro}, {Frenk} \& {White}}{{Navarro}
  et~al.}{1996}]{navarro96}
{Navarro} J.~F.,  {Frenk} C.~S.,    {White} S.~D.~M.,  1996, ApJ, 462, 563

\bibitem[\protect\citeauthoryear{{Ostriker} \& {Peebles}}{{Ostriker} \&
  {Peebles}}{1973}]{ostriker73}
{Ostriker} J.~P.,  {Peebles} P. J.~E.,  1973, ApJ, 186, 467

\bibitem[\protect\citeauthoryear{{Patsis}, {Grosb{\o}l} \& {Hiotelis}}{{Patsis}
  et~al.}{1997}]{patsis97a}
{Patsis} P.~A.,  {Grosb{\o}l} P.,    {Hiotelis} N.,  1997, A\&A, 323, 762

\bibitem[\protect\citeauthoryear{{Patsis}, {Hiotelis}, {Contopoulos} \&
  {Grosb{\o}l}}{{Patsis} et~al.}{1994}]{patsis94}
{Patsis} P.~A.,  {Hiotelis} N.,  {Contopoulos} G.,    {Grosb{\o}l} P.,  1994,
  A\&A, 286, 46

\bibitem[\protect\citeauthoryear{{Rautiainen} \& {Salo}}{{Rautiainen} \&
  {Salo}}{1999}]{rautiainen99}
{Rautiainen} P.,  {Salo} H.,  1999, A\&A, 348, 737

\bibitem[\protect\citeauthoryear{{Rautiainen} \& {Salo}}{{Rautiainen} \&
  {Salo}}{2000}]{rautiainen2000}
{Rautiainen} P.,  {Salo} H.,  2000, A\&A, 362, 465

\bibitem[\protect\citeauthoryear{{Rautiainen}, {Salo} \&
  {Laurikainen}}{{Rautiainen} et~al.}{2002}]{rautiainen2002}
{Rautiainen} P.,  {Salo} H.,    {Laurikainen} E.,  2002, MNRAS, 337, 1233

\bibitem[\protect\citeauthoryear{{Regan} \& {Teuben}}{{Regan} \&
  {Teuben}}{2003}]{regan2003}
{Regan} M.~W.,  {Teuben} P.,  2003, ApJ, 582, 723

\bibitem[\protect\citeauthoryear{{Rix} \& {Rieke}}{{Rix} \&
  {Rieke}}{1993}]{rix93}
{Rix} H.,  {Rieke} M.~J.,  1993, ApJ, 418, 123

\bibitem[\protect\citeauthoryear{{Ryder}, {Buta}, {Toledo}, {Shukla},
  {Staveley-Smith} \& {Walsh}}{{Ryder} et~al.}{1996}]{ryder96}
{Ryder} S.~D.,  {Buta} R.~J.,  {Toledo} H.,  {Shukla} H.,  {Staveley-Smith} L.,
     {Walsh} W.,  1996, ApJ, 460, 665

\bibitem[\protect\citeauthoryear{{Sackett}}{{Sackett}}{1997}]{sackett97}
{Sackett} P.~D.,  1997, ApJ, 483, 103

\bibitem[\protect\citeauthoryear{{Salo}, {Rautiainen}, {Buta}, {Purcell},
  {Cobb}, {Crocker} \& {Laurikainen}}{{Salo} et~al.}{1999}]{salo99}
{Salo} H.,  {Rautiainen} P.,  {Buta} R.,  {Purcell} G.~B.,  {Cobb} M.~L.,
  {Crocker} D.~A.,    {Laurikainen} E.,  1999, AJ, 117, 792

\bibitem[\protect\citeauthoryear{{Sandage}}{{Sandage}}{1961}]{sandage61}
{Sandage} A.,  1961, The Hubble atlas of galaxies.
Washington: Carnegie Institution, 1961

\bibitem[\protect\citeauthoryear{{Sandage} \& {Bedke}}{{Sandage} \&
  {Bedke}}{1994}]{sandage94}
{Sandage} A.,  {Bedke} J.,  1994, The Carnegie atlas of galaxies.
Washington, DC: Carnegie Institution of Washington with The Flintridge
  Foundation

\bibitem[\protect\citeauthoryear{{Schommer} \& {Sullivan}}{{Schommer} \&
  {Sullivan}}{1976}]{schommer76}
{Schommer} R.~A.,  {Sullivan} W.~T.,  1976, Astrophysical Letters, 17, 191

\bibitem[\protect\citeauthoryear{{Schwarz}}{{Schwarz}}{1981}]{schwarz81}
{Schwarz} M.~P.,  1981, ApJ, 247, 77

\bibitem[\protect\citeauthoryear{{Schwarz}}{{Schwarz}}{1984a}]{schwarz84b}
{Schwarz} M.~P.,  1984a, MNRAS, 209, 93

\bibitem[\protect\citeauthoryear{{Schwarz}}{{Schwarz}}{1984b}]{schwarz84a}
{Schwarz} M.~P.,  1984b, A\&A, 133, 222

\bibitem[\protect\citeauthoryear{{Sellwood}}{{Sellwood}}{1981}]{sellwood81}
{Sellwood} J.~A.,  1981, A\&A, 99, 362

\bibitem[\protect\citeauthoryear{{Sellwood} \& {Sparke}}{{Sellwood} \&
  {Sparke}}{1988}]{sellwood88}
{Sellwood} J.~A.,  {Sparke} L.~S.,  1988, MNRAS, 231, 25P

\bibitem[\protect\citeauthoryear{{Sempere}, {Garcia-Burillo}, {Combes} \&
  {Knapen}}{{Sempere} et~al.}{1995}]{sempere95a}
{Sempere} M.~J.,  {Garcia-Burillo} S.,  {Combes} F.,    {Knapen} J.~H.,  1995,
  A\&A, 296, 45

\bibitem[\protect\citeauthoryear{{Teuben}}{{Teuben}}{2002}]{teuben2002}
{Teuben} P.~J.,  2002, ASP Conf. Ser. 275: Disks of Galaxies: Kinematics,
  Dynamics and Perturbations, p.~217

\bibitem[\protect\citeauthoryear{{Tremaine} \& {Weinberg}}{{Tremaine} \&
  {Weinberg}}{1984}]{tremaine84}
{Tremaine} S.,  {Weinberg} M.~D.,  1984, ApJ, 282, L5

\bibitem[\protect\citeauthoryear{{Valenzuela} \& {Klypin}}{{Valenzuela} \&
  {Klypin}}{2002}]{valenzuela2002}
{Valenzuela} O.,  {Klypin} A.,  2002, astro-ph/0204028

\bibitem[\protect\citeauthoryear{{van Albada} \& {Sanders}}{{van Albada} \&
  {Sanders}}{1982}]{vanalbada82}
{van Albada} T.~S.,  {Sanders} R.~H.,  1982, MNRAS, 201, 303

\bibitem[\protect\citeauthoryear{{Warner}, {Wright} \& {Baldwin}}{{Warner}
  et~al.}{1973}]{warner73}
{Warner} P.~J.,  {Wright} M. C.~H.,    {Baldwin} J.~E.,  1973, MNRAS, 163, 163

\bibitem[\protect\citeauthoryear{{Weinberg}}{{Weinberg}}{1985}]{weinberg85}
{Weinberg} M.~D.,  1985, MNRAS, 213, 451

\bibitem[\protect\citeauthoryear{{Weiner}, {Sellwood} \& {Williams}}{{Weiner}
  et~al.}{2001}]{weiner2001b}
{Weiner} B.~J.,  {Sellwood} J.~A.,    {Williams} T.~B.,  2001, ApJ, 546, 931

\end{thebibliography}

\end{document}